\documentclass[%
aps,%
prd,%
reprint,%
preprintnumbers,%
showpacs,%
amsmath,%
amssymb,%
nofootinbib,%
superscriptaddress,%
twoside%
]{revtex4-1}

\usepackage{amsfonts}
\usepackage{graphicx}
\usepackage{xcolor}
\usepackage{mathtools}
\usepackage{bm}
\usepackage{dsfont}
\usepackage{hyperref}
\usepackage{braket}

\newcommand{\eq}[1]{\begin{align} #1 \end{align}}

\newcommand{\del}{\partial}
\newcommand{\e}{\mathrm{e}}
\newcommand{\Tr}{\operatorname{Tr}}
\newcommand{\tr}{\operatorname{tr}}
\newcommand{\diff}{\mathrm{d}}

\newcommand{\config}{\mathcal{F}}
\newcommand{\gauge}{\mathcal{G}}
\newcommand{\vp}{\varphi}
\newcommand{\vpb}{\bar{\varphi}}

\newcommand{\gb}{\bar{g}}
\newcommand{\bg}{\bar{g}}
\newcommand{\mat}{\mathcal{M}}
\newcommand{\matrices}{\mathbb{R}^{d\times d}}
\newcommand{\mn}{{\mu\nu}}
\newcommand{\GLd}{\operatorname{GL}(d)}
\newcommand{\Opq}{\operatorname{O}}
\newcommand{\bo}{{\bar{o}}}
\newcommand{\mku}{\mkern1mu}
\newcommand{\fg}{\mathfrak{g}}
\newcommand{\fh}{\mathfrak{h}}
\newcommand{\fm}{\mathfrak{m}}
\newcommand{\Ad}{\mathrm{Ad}}
\newcommand{\calD}{\mathcal{D}}
\newcommand{\frakD}{\mathfrak{D}}
\newcommand{\mO}{\mathcal{O}}
\newcommand{\bG}{\bar{\Gamma}}
\newcommand{\bp}{\bar{\varphi}}
\newcommand{\sg}{\sqrt{g}}
\newcommand{\sbg}{\sqrt{\bar{g}}}

\begin{document}

\title{Connections and geodesics in the space of metrics}

\author{Maximilian Demmel}
\email{demmel@thep.physik.uni-mainz.de}
\affiliation{PRISMA Cluster of Excellence, Institute of Physics, Johannes Gutenberg University Mainz,
Staudingerweg 7, 55099 Mainz, Germany}
\affiliation{
Institute for Mathematics, Astrophysics and Particle Physics (IMAPP),\\
Radboud University Nijmegen, Heyendaalseweg 135, 6525 AJ Nijmegen, The Netherlands	
}

\author{Andreas Nink}
\email{nink@thep.physik.uni-mainz.de}
\affiliation{PRISMA Cluster of Excellence, Institute of Physics, Johannes Gutenberg University Mainz,
Staudingerweg 7, 55099 Mainz, Germany}

\begin{abstract}
We argue that the exponential relation $g_\mn = \bg_{\mu\rho}\big(\e^h\big)^\rho{}_\nu$ is the most natural
metric parametrization since it describes geodesics that follow from the basic structure of the space of metrics.
The corresponding connection is derived, and its relation to the Levi-Civita connection and the
Vilkovisky-DeWitt connection is discussed. We address the impact of this geometric formalism on quantum gravity
applications. In particular, the exponential parametrization is appropriate for constructing covariant quantities
like a reparametrization invariant effective action in a straightforward way. Furthermore, we reveal an important
difference between Euclidean and Lorentzian signatures: Based on the derived connection, any two Euclidean
metrics can be connected by a geodesic, while this does not hold for the Lorentzian case.
\end{abstract}

\preprint{MITP/15-041}

\maketitle
\section{Introduction}
A metric on a manifold is a covariant rank-$2$ tensor field (i.e.\ it is continuous and bi-linear) which is
symmetric and non-degenerate. As a consequence of continuity and non-degeneracy, the signature of any metric is
constant. Fixing the signature restricts the set of all possible metrics. When we speak about ``space of
metrics'' in this article we assume that all elements have the same prescribed signature. It is one aim to
discuss the fundamentals of the geometry of such a space. In fact, the requirement for non-degeneracy, or
equivalently, for a fixed signature, imposes a \emph{non-linear constraint} on metrics, and all non-trivial
geometric properties of the space of metrics are due to this requirement.

While the constraint must be strictly satisfied in General Relativity, it is not clear a priori if it should be
respected in a gravitational path integral, too \cite{Percacci:1991}. However, in the class of actions that we
would like to consider, i.e.\ functionals constructed from invariants of the type $\int\diff^d x\sg$,
$\int\diff^d x\sg R$, etc., it is crucial to have a non-degenerate metric. Otherwise, the volume element
$\sg$ could vanish and the inverse metric required to raise indices could be non-existent. Therefore, we take
the view that the constraint has also to be taken into account in the domain of integration in a path integral.

The application of conventional quantum field theory methods to gravity requires the
introduction of a background metric, say, $\gb_\mn$. Usually, the dynamical metric $g_\mn$ is split into
background and fluctuations $h_\mn$ in the standard linear way by writing
\begin{equation}
 g_\mn = \gb_\mn + h_\mn.
\label{eqn:StdParam}
\end{equation}
Due to the non-linear constraint, however, the space of metrics is not a vector space, and thus, the addition
in equation \eqref{eqn:StdParam} has to be handled with care. There are two ways to approach this difficulty
\cite{DeWitt:2003}. (a) If $\bg_\mn$ and $g_\mn$ lie in the same coordinate patch, then $h_\mn$ can simply be seen
as a coordinate increment, where the addition is well defined in the chart. (b) In general, one should regard
$h_\mn$ as components of a tangent vector to the space of metrics at $\bg$. Then the ``addition'' in
\eqref{eqn:StdParam} is to be understood as starting at the base point $\bg_\mn$ and going along a geodesic in the
direction of $h_\mn$ which assumes the role of geodesic (normal) coordinates. As we will see, both points of view
ultimately lead to a more natural parametrization of metrics in comparison with the linear one.

The gravitational path integral is given by an integration over the metric fluctuations, $\int\calD h_\mn$.
Now the different notions (a) and (b), which we refer to as non-geometric and geometric, respectively, lead to
different implementations of the constraint for the metrics. In the non-geometric interpretation (a) we have to
\emph{restrict the domain of integration} to the ($\gb_\mn$-dependent) subset of those $h_\mn$ which define
allowed metrics when using \eqref{eqn:StdParam}. This can be done by reparametrizing the metric such that it
automatically has the correct signature for all fluctuations and that the new domain of integration is trivial.
By contrast, in case (b) the constraint is \emph{satisfied already by construction}. The fluctuations are
interpreted as tangent vectors that are inserted into the exponential map of the space of metrics, thus giving
rise to admissible metrics only.

Motivated by a different argument\footnote{The exponential parametrization allows for an easy separation of the
conformal mode from the fluctuations.}, one particular metric parametrization has been used previously in
reference \cite{Kawai:1993}. It is given by the exponential relation
\begin{equation}
 g_\mn = \bg_{\mu\rho}\big(\e^h\big)^\rho{}_\nu \, ,
\label{eqn:ExpParam}
\end{equation}
where indices are raised and lowered with the background metric, and $h$ is a symmetric matrix-valued field,
$h_\mn = h_{\nu\mu}$ (or $h^\mu{}_\nu= h_\nu{}^\mu$ with the shifted index position). In matrix notation
equation \eqref{eqn:ExpParam} reads
\begin{equation}
 g = \bg\,\e^{\bg^{-1}h},
\label{eqn:ExpParamMatrix}
\end{equation}
with $h^T=h$. Note that the factor $\bg^{-1}$ in the exponent is meant implicitly in equation
\eqref{eqn:ExpParam}, indicated by the index positions. In our present context, the special significance of this
exponential parametrization lies in the fact that it satisfies the above constraint.

To see this, let us first adopt the non-geometric interpretation. From that point of view we regard equation
\eqref{eqn:ExpParam} as a mere \emph{change of coordinates} from metrics $g_\mn$ to symmetric tensors $h_\mn$,
i.e.\ as a \emph{reparametrization}. It can be shown without relying on geometric constructions based on geodesics
that this is a one-to-one correspondence \cite{Nink:2014yya}.\footnote{The one-to-one correspondence was shown for
Euclidean metrics. For Lorentzian signatures see section \ref{sec:EuLor}.} That is, not only does the right hand
side of \eqref{eqn:ExpParam} give rise to admissible metrics, but there also exists a unique symmetric $h_\mn$ for
any given $g_\mn$ and $\bg_\mn$. Hence, the required signature constraint is satisfied, and the path integral over
the fluctuations $h_\mn$ captures every $g_\mn$ once and only once.

On the other hand, let us take the geometric view now, where $h_\mn$ assumes the role of a \emph{tangent
vector}. In the remainder of this article we will always take this view, unless stated otherwise. It allows for
profound insights into the structure of the space of metrics. Remarkably enough, it leads to the same parametrization
\eqref{eqn:ExpParam} as above. This is worked out in detail in section \ref{sec:GroupTheory} (cf.\ also
\cite{DeWitt:1967,Freed:1989,Percacci:2015wwa}). The construction is based on \emph{geodesics}, and thus, the
parametrization clearly depends on the underlying \emph{connection}. We will argue, however, that there is one
natural choice of a connection which results from the basic properties of metrics. Accordingly, we consider the
exponential relation \eqref{eqn:ExpParam} the most natural parametrization.

Apart from its fundamental geometric meaning and its advantage of generating only such metrics that satisfy
the signature constraint, the exponential parametrization is further motivated by several physical arguments.
Here we briefly mention some of them.
\renewcommand{\theenumi}{(\roman{enumi})}
\begin{enumerate}
 \item As already indicated above, the use of parametrization \eqref{eqn:ExpParam} allows for an easy separation
  of the conformal mode from the fluctuations: When splitting $h_\mn$ into trace and traceless contributions,
  $h_\mn=\hat{h}_\mn+\frac{1}{d}\bg_\mn \phi$, with $\phi=\bg^\mn h_\mn$ and $\bg^\mn\hat{h}_\mn=0$, the trace
  part gives rise to a conformal factor in \eqref{eqn:ExpParam}, and notably, the volume element on the spacetime
  manifold depends only on $\phi$, $\sg=\sbg\, \e^{\frac{1}{2}\phi}$. In the context of gravity this means that
  the cosmological constant occurs as a coupling only in the conformal mode sector. This is one reason for the
  following point.
\item Some computations are simplified and some are feasible only when using parametrization \eqref{eqn:ExpParam},
  for instance to avoid infrared singularities in the search of scaling solutions in scalar-tensor gravity
  \cite{Percacci:2015wwa,Labus:2015ska}, for calculating the limit $\epsilon\rightarrow 0$ of the effective action
  in $2+\epsilon$ dimensional quantum gravity \cite{Codello:2014}, in unimodular quantum gravity
  \cite{Eichhorn:2013}, and for ensuring gauge independence at one loop level without resorting to the
  Vilkovisky-DeWitt method \cite{Falls:2015} (cf.\ also section \ref{sec:CompConn}).
\item It is known from conformal field theory studies that there is a critical number of scalar fields
  in a theory of gravity coupled to conformal matter, referred to as the critical central charge, which
  amounts to $c_\text{crit}=25$ \cite{David:1988}. This result is correctly reproduced in the Asymptotic
  Safety program \cite{Weinberg:1980} when using the exponential parametrization
  \cite{Kawai:1993,Nink:2014yya,Codello:2014}, while a different number is obtained when using the linear
  relation \eqref{eqn:StdParam} \cite{Tsao:1977,Nink:2014yya,Codello:2014}.
\end{enumerate}

At this point a comment is in order. The equivalence theorem \cite{Borchers:1960} states invariance of the
$S$-matrix, and thus, of all physical quantities, under field redefinitions. With this in mind let us discuss
why the choice of parametrization matters at all.

The first point we want to make is that the linear split \eqref{eqn:StdParam} is often taken seriously where
the addition is the usual tensor addition. The path integral is then thought of as an integration over all
symmetric tensors. This way, it would be easy to evaluate Gaussian integrals \cite{Mottola:1995}, for instance.
However, as discussed above, metrics have to satisfy the signature constraint which amounts to a restricted
domain of integration. Therefore, \emph{the exponential parametrization} \eqref{eqn:ExpParam} \emph{is a field
redefinition of} \eqref{eqn:StdParam} \emph{only if the latter is combined with the constraint}. Only then the
$S$-matrices can be expected to agree.

Secondly, one is often interested in off shell quantities, e.g.\ in $\beta$-functions for renormalization
group studies or in the effective potential part of the effective action for investigating spontaneous
symmetry breaking. In general, off shell quantities depend on the choice of parametrization. This fact can
be important when comparing different approaches that describe the same physics. For instance, there are
several candidate theories of quantum gravity, and the use of a particular parametrization in one theory
might be most appropriate for a comparison with another one. So the choice of parametrization can indeed be
relevant in the usual non-invariant framework, and it can be a powerful tool to simplify computations.

Pioneered by Vilkovisky \cite{Vilkovisky:1984st} and DeWitt \cite{DeWitt:1987}, there is, however, a way to
construct an effective action $\Gamma$ which is reparametrization invariant and gauge independent both off and
on shell. The price one has to pay for this invariance is a nontrivial dependence of $\Gamma$ on the background
metric, encoded in generalized Nielsen identities \cite{Burgess:1987} (cf.\ section \ref{sec:Applications}), which
obscures relations between variables and makes calculations more complex. As we will argue, \emph{the geometric
interpretation of} \eqref{eqn:ExpParam} \emph{leads to reparametrization and gauge invariant (but not gauge
independent) constructions}, too, and it entails a simpler relation between two metrics connected by a geodesic
as compared to the Vilkovisky-DeWitt approach. Thus, it depends on the desired application whether a
reparametrization invariant method is useful, and which connection for determining geodesics should be chosen.

Above we have seen the significance of the exponential parametrization with its geometric
meaning and its many advantages for physical applications. The present work is dedicated to
investigating the geometric structure behind it. We aim at finding a connection in field space such that the
corresponding geodesics are parametrized by relation \eqref{eqn:ExpParam}. To put it another way, we
determine \emph{a connection such that the exponential map is given by the standard matrix exponential}.

Some of the arguments brought up here for our calculations are already known. Our objective is to collect them,
supplement them further, compare different approaches and embed the ideas into a broader context. This article is
organized as follows. In section \ref{sec:DerivationNDConnection} we present a derivation of a connection that
leads to parametrization \eqref{eqn:ExpParam}. We compare this connection with the Levi-Civita connection and the
Vilkovisky-DeWitt connection in section \ref{sec:CompConn}, starting from a metric in field space. The main
part is contained in section \ref{sec:GroupTheory}: We rederive the connection of section
\ref{sec:DerivationNDConnection} with more general methods borrowed from group theory and differential
geometry, where we find that it originates from a basic geometric structure that is given in a natural way.
Furthermore, we study differences between the space of Euclidean and Lorentzian metrics, see section
\ref{sec:EuLor}. In section \ref{sec:Applications} we discuss the meaning of the exponential
parametrization for its application to covariant Taylor expansions and Nielsen identities. Finally,
we conclude with a short summary in section \ref{sec:Conclusion}.

\section{Derivation of the connection}
\label{sec:DerivationNDConnection}
Geodesics on a differentiable manifold -- parametrized by means of the \emph{exponential map} -- are fixed by the
choice of an affine connection. In this context, different connections lead to different exponential maps. Since we
have already discussed the importance of the metric parametrization \eqref{eqn:ExpParam}, we now aim at finding a
connection on the space of metrics such that the exponential map has the simple form of the standard \emph{matrix
exponential}.

Before we start, let us briefly fix the notations and conventions used in this article. The spacetime manifold
is denoted by $M$, and points in $M$ by $x,y,z$. The set of all field configurations is referred to as \emph{field
space}, henceforth denoted by $\config$. In the present case, $\config$ is the space of all metrics on $M$. It can
be shown that $\config$ exhibits the structure of an (infinite dimensional) manifold
\cite{Ebin:1970,Gil-Medrano:1991,Blair:2000}. We observe that any spacetime metric $g\in\config$ at a given spacetime
point can be considered a symmetric matrix. More precisely, if $g$ has signature $(p,q)$, then in any chart $(U,\phi)$
for the spacetime manifold $M$ the metric in local coordinates is a map
\begin{equation}
\label{eq:mapping}
 g\big|_U:U\rightarrow \mat \, , \quad x\mapsto g_\mn(x),
\end{equation}
where $\mat$ denotes the set of real non-degenerate symmetric $d\times d$ matrices with signature\footnote{In our
convention, $p$ is the number of positive eigenvalues and $q$ the number of negative ones. Due to non-degeneracy we
have $p+q=d$. Matrices with $p=d$, $q=0$ are positive definite, corresponding to Euclidean metrics.} $(p,q)$,
\begin{equation}
 \mat \equiv \left\{ A \in \GLd \big| \, A^T=A,\; A \text{ has signature }(p,q)\right\}.
\label{eqn:DefMatrices}
\end{equation}
Due to this local appearance of metrics at a given point, we may think of the configuration space $\config$ as the
topological product $\prod_{x\in M} \mat$. In practice, this notion has to be supplemented by additional requirements
concerning continuity. Actually, $\config$ is the space of sections of a fiber bundle with typical fiber $\mat$ and
base space $M$, but in the present context it is not necessary to specify this further. As we will argue, geodesics in
$\config$ are closely related to geodesics in $\mat$ for a certain class of connections.

A generic field $\vp^i$ can be regarded as the local coordinate representation of a point in field space $\config$.
We employ DeWitt's condensed notation \cite{DeWitt:1965jb}, where the (Latin) index $i$ represents both discrete and
continuous (e.g.\ spacetime) labels, so we identify $\vp^i\equiv g_\mn(x)$. Repeated condensed indices are interpreted
as summation over discrete and integration over continuous indices. By $\bp^i$ we denote a fixed but arbitrary
background field.

Our starting point for the derivation of the desired connection will be an expansion of $\vp^i$ in terms of tangent
vectors of a geodesic connecting $\bp^i$ and $\vp^i$. Let $\vp^i(s)$ denote such a geodesic, i.e.\ a curve with
\begin{equation}
\vp^i(0)=\bp^i\quad \text{and}\quad \vp^i(1)=\vp^i,
\end{equation}
that satisfies the geodesic equation
\begin{equation}
	\ddot{\vp}^i(s)+\Gamma^i_{jk}\,\dot{\vp}^j(s)\dot{\vp}^k(s)=0,
\label{eqn:GeodEqu}
\end{equation}
where the dots indicate derivatives w.r.t.\ the curve parameter $s$, and $\Gamma^i_{jk}$ is the Christoffel symbol
evaluated at $\vp^i(s)$, i.e.\ $\Gamma^i_{jk}\equiv\Gamma^i_{jk}[\vp^i(s)]$. We assume for a moment that the geodesic
$\vp^i(s)$ lies entirely in one coordinate patch. As we will see, the connection determined below only gives rise to
such geodesics that automatically satisfy this assumption. Then we can expand the local coordinates as a series,
\begin{equation}
	\vp^i(s)=\sum\limits_{n=0}^\infty\frac{s^n}{n!}\left(\frac{\diff^n}{\diff s^n}\vp^i(s)\Big|_{s=0}\right).
\label{eqn:PhiExpansion0}
\end{equation}
We observe that it is possible to express all higher derivatives in \eqref{eqn:PhiExpansion0} in terms of
$\dot{\vp}^i$ by using equation \eqref{eqn:GeodEqu} iteratively. If $h^i\equiv\dot{\vp}^i(0)$ denotes the tangent
vector at $\bp$ in the direction of the geodesic, we obtain the following relation for $\vp^i=\vp^i(1)$:
\begin{equation}
\begin{split}
  \vp^i={}&\bp^i+h^i -{\textstyle\frac{1}{2}}\mku\bG^i_{jk}\,h^j h^k\\
    &+{\textstyle\frac{1}{6}}\big(\bG^i_{mj}\bG^m_{lk}+\bG^i_{km}\bG^m_{lj}-\bG^i_{jk,l}\big)h^j h^k h^l +\mO(h^4),
\end{split}
\label{eqn:PhiExpansion}
\end{equation}
with $\bG^i_{jk}=\Gamma^i_{jk}[\bp]$ and $\bG^i_{jk,l}\equiv \frac{\delta}{\delta\bp^l}\bG^i_{jk}$. In standard
index notation equation \eqref{eqn:PhiExpansion} reads
\begin{equation}
\begin{split}
	g_\mn(x)={}&\bg_\mn(x)+h_\mn(x)\\ &-{\textstyle\frac{1}{2}}\int_y\int_z
		\bG^{\alpha\beta\,\rho\sigma}_\mn(x,y,z)h_{\alpha\beta}(y) h_{\rho\sigma}(z) +\mO(h^3).
\end{split}
\label{eqn:gExpansion}
\end{equation}
This expansion is to be compared with the exponential metric parametrization \eqref{eqn:ExpParam}, which can be
written as the series
\begin{equation}
  g_\mn(x) = \bg_\mn(x)+h_\mn(x)+{\textstyle\frac{1}{2}}\mku\bg^{\rho\sigma}(x)h_{\mu\rho}(x)h_{\nu\sigma}(x) +\mO(h^3).
\label{eqn:gExpSeries}
\end{equation}
From the second order terms in \eqref{eqn:gExpansion} and \eqref{eqn:gExpSeries} we can finally read off the
connection $\bG^{\alpha\beta\,\rho\sigma}_\mn(x,y,z)$.\footnote{Since we must take into account that the affine
connection maps again to an element of the tangent space, i.e.\ to a symmetric tensor, we have to symmetrize
adequately. By convention, round brackets indicate symmetrization:
$a_{(\mn)}\equiv\frac{1}{2}(a_{\mn}+a_{\nu\mu})$.} Since the result is valid for arbitrary base points
$\bg_\mn$, we can go over to its unbarred version, i.e.\ to the connection evaluated at $g_\mn$, and we obtain
\begin{equation}
  \Gamma^{\alpha\beta\,\rho\sigma}_\mn(x,y,z)=-\delta^{(\alpha}_{(\mu}\,
  g^{\raisebox{0.2ex}{$\scriptstyle\beta)(\rho$}}(x)\,
  \delta^{\sigma)}_{\nu)}\; \delta(x-y)\delta(x-z).
\label{eqn:NDConnection}
\end{equation}
This is the main result of this section.

It remains to be shown that the connection \eqref{eqn:NDConnection} is consistent also with all higher orders in
\eqref{eqn:gExpansion} and \eqref{eqn:gExpSeries}. One can check as an easy exercise that the third order terms
do in fact agree. For a proof at all orders, however, we proceed differently. The idea is to find exact solutions
to the geodesic equation \eqref{eqn:GeodEqu} based on the connection \eqref{eqn:NDConnection}.

But first, we make an important remark about a fundamental property of the connection. Since
$\Gamma^{\alpha\beta\,\rho\sigma}_\mn(x,y,z)$ is proportional to $\delta(x-y)\delta(x-z)$, all integrations in
\eqref{eqn:GeodEqu} are trivial. Thus, the geodesic equation is \emph{effectively pointwise}
with respect to the spacetime. As already stated above, at any given point $x$ the metric can be considered an element
of $\mat$, defined in \eqref{eqn:DefMatrices}, which is an open and connected subset in the vector space of symmetric
matrices (cf.\ discussion in section \ref{sec:GroupTheory}), and which can thus be covered with one coordinate chart.
Therefore, geodesics corresponding to \eqref{eqn:NDConnection} stay indeed in one chart.

Due to the pointwise character of the geodesic equation, the dependence on $x$ is not written explicitly in the
following. Now equation \eqref{eqn:GeodEqu} becomes
\begin{equation}
	\ddot{g}_\mn -\delta^{(\alpha}_{(\mu}\, g^{\raisebox{0.2ex}{$\scriptstyle\beta)(\rho$}}\,\delta^{\sigma)}_{\nu)}
	\dot{g}_{\alpha\beta} \dot{g}_{\rho\sigma}
	= \ddot{g}_\mn -g^{\beta\rho}\dot{g}_{\mu\beta} \dot{g}_{\rho\nu}=0.
\label{eqn:NewGeodEqu}
\end{equation}
After multiplication with $g^{\nu\lambda}$, we observe that \eqref{eqn:NewGeodEqu} can be brought to the form
\begin{equation}
\frac{\diff}{\diff s}\left(\dot{g}_\mn g^{\nu\lambda}\right)=0,
\end{equation}
that is, $\dot{g}_\mn g^{\nu\lambda}=c^\lambda_\mu=\text{const}$. In matrix notation this reads
\begin{equation}
	\dot{g}(s)=c\mku g(s).
\label{eqn:MatDiffEq}
\end{equation}
Equation \eqref{eqn:MatDiffEq} is known to have the unique solution $g(s)=\e^{sc}g(0)$. With the initial conditions
$g(0)=\bg$ and $h=\dot{g}(0)=c\mku g(0)=c\mku\bg$ we obtain $g(s)=\e^{s\mku h\bg^{-1}} \bg$, which
finally leads to
\begin{equation}
	g(s)=\bg\,\e^{s\mku \bg^{-1} h}.
\label{eqn:GeodInM}
\end{equation}
At $s=1$ and in index notation this is precisely the exponential relation \eqref{eqn:ExpParam} for the metric.
Hence we have proven that geodesics corresponding to the connection \eqref{eqn:NDConnection} are uniquely
parametrized by $g_\mn = \bg_{\mu\rho}\big(\e^h\big)^\rho{}_\nu$. As a result, \eqref{eqn:gExpansion} and
\eqref{eqn:gExpSeries} agree at all orders. Note that equation \eqref{eqn:GeodInM} defines a geodesic in
$\mat$, too, as it holds at each spacetime point $x$ separately, while it becomes a geodesic in $\config$
when regarding $\bg$ and $h$ as $x$-dependent tensor fields. Continuity of $g$ with respect to $x$ is then
ensured by continuity of $\bg$ and $h$.

In conclusion, there is indeed a connection that defines a structure on field space $\config$ entailing a
simple parametrization of geodesics. Whether there is even more structure by virtue of a field space
metric will be discussed in the following section.

\section{Comparison of connections on field space}
\label{sec:CompConn}

Above we showed that the field space $\config$ can be equipped with a connection $\Gamma^k_{ij}$ that
reproduces the exponential parametrization. Now, we discuss different connections on field space known from
the literature and their relation to the new connection \eqref{eqn:NDConnection}.

As we already described in the previous section, the metric $g_\mn$ is a map from the spacetime manifold $M$ to
the set of non-degenerate symmetric matrices $\mat$, which by itself carries the structure of a manifold.
Including field space, we are dealing with three manifolds in total, which we carefully distinguish. We will
see that all of them can be equipped with a metric, leading to the three (semi-) Riemannian manifolds
\begin{align}
(M,g), \qquad (\mat,\gamma),\qquad (\config,G)\,,
\end{align}
where $g_{\mu\nu}$ is the spacetime metric, $\gamma$ is the metric in $\mat$ and $G_{ij}$ denotes the field
space metric. Note that $g_\mn$ also represents a point in $\config$. The field space metric $G_{ij}$ is part
of the definition of the theory under consideration, but nevertheless, it can be fixed if a few requirements
are made.

Firstly, we want to take into account that gravity is a gauge theory. The classical action is invariant under
diffeomorphisms, and so are all physical quantities. This leads to the reasonable requirement that the metric
$G_{ij}$ on $\config$ be gauge invariant, too, i.e.\ that the action of the gauge group on $\config$ be an
isometry. In general terms, a gauge transformation can be written as
\eq{
  \label{eqn:GaugeTrafo}
  \delta \vp^i = K_\alpha^i[\vp] \delta\epsilon^\alpha\, ,
}
where $\delta\epsilon^\alpha$ parametrizes the transformation and the $\mathbf{K}_\alpha$ are the generators
of the gauge group $\gauge$. In the case of gravity, equation \eqref{eqn:GaugeTrafo} reads
$\delta g_\mn=\mathcal{L}_{\delta\epsilon}g_\mn$, with the Lie derivative $\mathcal{L}$ along a vector field
$\delta \epsilon^\alpha$. The action of $\gauge$ on $\config$ induces a principal bundle structure.
Points that are connected by gauge transformations are physically equivalent while the space of orbits
$\config/\gauge$ contains all physically nonequivalent configurations. Now, if the gauge group is to generate
isometric motions in $\config$, then the field space metric $G_{ij}[\vp]$ must satisfy Killing's equation,
i.e.\ our first requirement reads
\begin{equation}
K^k_{\alpha,i} G_{jk} + K^k_{\alpha,j} G_{ik} + K^k_{\alpha} G_{ij,k} =0\,,
\end{equation}
where commas denote functional derivatives with respect to the field $\vp^i$.

Secondly, we require that $G_{ij}[\vp]$ be ultra-local, i.e. that it involve only undifferentiated $\vp$'s and
that it be diagonal in $x$-space.

There is a unique one-parameter family of field space metrics satisfying all requirements, which is known as
DeWitt metric \cite{DeWitt:1967}. It reads
\begin{equation}
  G^{\mu\nu\,\rho\sigma}(x,y)[g] = \sg \left(g^{\mu(\rho} g^{\sigma)\nu} + \frac{c}{2}\, g^{\mu\nu}
  g^{\rho\sigma}\right) \delta(x-y) \,,
\label{eqn:DeWitt_metric}
\end{equation}
where the $x$-dependence of $g_\mn$ is implicit. This metric on $\config$ is our starting point.

From it we can deduce a metric on $\mat$ as well by identifying it with the tensor part of the DeWitt metric.
(The factor $\sg$ in \eqref{eqn:DeWitt_metric} is needed only to make $G^{\mu\nu\,\rho\sigma}(x,y)$ a bi-tensor
density of correct weight.) That is, we define
\begin{equation}
 \gamma^{\mu\nu\,\rho\sigma}(g) \equiv g^{\mu(\rho} g^{\sigma)\nu} + \frac{c}{2}\, g^{\mu\nu} g^{\rho\sigma} \,.
\label{eqn:NDMetric}
\end{equation}
Hence, the DeWitt metric can be written as
\begin{equation}
  G^{\mu\nu\,\rho\sigma}(x,y)[g] = \sqrt{g(x)}\, \gamma^{\mu\nu\,\rho\sigma} (g(x))\, \delta(x-y) \,.
\label{eqn:ultra_local_metric}
\end{equation}

Next, we determine the Levi-Civita (LC) connection on $\mat$ w.r.t.\ the metric \eqref{eqn:NDMetric},
where we point out the difference compared with the LC connection on $\config$ induced by the DeWitt metric.
In the following, capital Latin indices abbreviate pairs of spacetime indices, e.g.\ $g^I(x) \equiv
g_{\mu\nu}(x)$. Let $\left\{ {}^{K}_{IJ} \right\}$ denote the LC connection on $\mat$. By definition we have
\begin{equation}
	\left\{ {}^K_{IJ} \right\} = \frac{1}{2}\gamma^{KL}\left(\gamma_{IL,J}+\gamma_{JL,I}-\gamma_{IJ,L}\right)\,.
\label{eqn:Levi-CivitaOnMDef}
\end{equation}
Notably, a direct calculation yields
\begin{equation}
	\left\{ {}^K_{IJ} \right\} \equiv \left\{ {}^{\alpha\beta\,\rho\sigma}_{\mu\nu} \right\}
	= -\delta^{(\alpha}_{(\mu}\, g^{\raisebox{0.2ex}{$\scriptstyle\beta)(\rho$}}\,\delta^{\sigma)}_{\nu)} \,,
\label{eqn:Levi-CivitaOnM}
\end{equation}
which has exactly the same tensor structure as our connection given by \eqref{eqn:NDConnection},
reproducing the exponential parametrization.

With this in mind, let us construct connections on field space $\config$ now. For that purpose we start from the
LC connection w.r.t.\ the DeWitt metric \eqref{eqn:DeWitt_metric}. It is denoted by $\left\{ {}^k_{ij} \right\}$,
and it follows from the usual definition,
\begin{equation}
	\left\{ {}^k_{ij} \right\} = \frac{1}{2} G^{kl} \left(G_{il,j}+ G_{jl,i} - G_{ij,l}\right) \,.
\label{eqn:Levi-Civita}
\end{equation}
Its form in terms of field space coordinates $g_\mn$ will be specified below. Now, a generic connection on
$\config$ can be written as 
\begin{equation}
	\Gamma_{ij}^k = \left\{ {}^k_{ij} \right\} + A_{ij}^k\,.
\label{eqn:Christoffels}
\end{equation}
The last term in \eqref{eqn:Christoffels} is an arbitrary smooth bi-linear bundle homomorphism, and different
connections on $\config$ merely differ in that term.

We would like to emphasize that, although by equation \eqref{eqn:ultra_local_metric} $G^{\mu\nu\,\rho\sigma}(x,y)$
is proportional to $\gamma^{\mu\nu\,\rho\sigma}$, the corresponding LC connections are not. The field space LC
connection rather contains additional terms. We find that it decomposes into two pieces,
\begin{equation}
  \left\{ {}^k_{ij} \right\} = \left(\left\{ {}^K_{IJ} \right\} + T^K_{IJ}\right)\!(x)\,\delta(x-y)\delta(x-z)\,,
\label{eqn:LCComparison}
\end{equation}
where the first term is given by equation \eqref{eqn:Levi-CivitaOnM} with $g_\mn$ replaced by $g_\mn(x)$, and
$T^K_{IJ}\equiv T_\mn^{\alpha\beta\,\rho\sigma}$ reads \cite{DeWitt:1967,Huggins:1987zw}
\begin{equation}
\begin{split}
 T_\mn^{\alpha\beta\,\rho\sigma} = \quad&\frac{1}{4} g^{\alpha\beta}\delta^\rho_{(\mu}\delta^\sigma_{\nu)}
  - \frac{1}{2(2+dc)} g_\mn g^{\alpha(\rho} g^{\sigma)\beta}\\
  + &\frac{1}{4} g^{\rho\sigma}\delta^\alpha_{(\mu}\delta^\beta_{\nu)}
  - \frac{c}{4(2+dc)} g_\mn g^{\alpha\beta} g^{\rho\sigma} \,.
\end{split}
\end{equation}
Clearly, the reason for this difference between the LC connections on $\mat$ and $\config$ can be traced to a
\emph{non-constant proportionality factor} relating the underlying metrics, i.e.\ to the volume element $\sqrt{g}$
in \eqref{eqn:ultra_local_metric}.
When taking functional derivatives of $G_{ij}$ they act both on $\sqrt{g}$ and on $\gamma^{\mu\nu\,\rho\sigma}$
in \eqref{eqn:ultra_local_metric}. Thus, the second term in \eqref{eqn:LCComparison} contains only contributions
due to derivatives acting on the volume element. This is a special characteristic of gravity. In other theories,
like in non-linear sigma models for instance \cite{Friedan:1980}, proportionality of a field space metric to a
metric in (the equivalent of) $\mat$ results in proportional LC connections. There the volume element is a
prescribed external ingredient, while it depends on the field in the case of gravity.

If we want to lift geodesics w.r.t.\ \eqref{eqn:Levi-CivitaOnM} from $\mat$ to $\config$, or, in other words,
if we want to obtain the connection \eqref{eqn:NDConnection} on $\config$ that reproduces the exponential
parametrization, we simply have to remove the terms originating from the volume element. This can easily be
achieved by choosing a bundle homomorphism $A_{ij}^k$ in \eqref{eqn:Christoffels} which takes the form
\begin{equation}
	A_{ij}^k = -T^K_{IJ}\,\delta(x-y)\delta(x-z) \,.
\label{eqn:AForND}
\end{equation}
That choice is perfectly admissible: All terms in $T^K_{IJ}$ are properly symmetrized, and thus, it maps two
symmetric tensors to a symmetric tensor again. Therefore, $A_{ij}^k$ represents a valid bundle homomorphism.
That way, we can indeed reconstruct our connection \eqref{eqn:NDConnection}.

For comparison, we would like to mention another famous choice for $A_{ij}^k$ which is due to Vilkovisky
\cite{Vilkovisky:1984st} and DeWitt \cite{DeWitt:1987}. It is adapted to the principal bundle structure of
$\config$ induced by the gauge group. The basic idea is to define geodesics on the physical base space
$\config/\gauge$ of the bundle and horizontally lift them to the full space $\config$. In this manner,
coordinates in field space are decomposed into gauge and gauge-invariant coordinates. The resulting
Vilkovisky-DeWitt connection is obtained by using \eqref{eqn:Christoffels} with the bundle homomorphism
\begin{equation}
	A_{ij}^k = K^\alpha_{(i} K^\beta_{j)} K^l_\alpha K^k_{\beta;l}-K^\alpha_i K^k_{\alpha;j}
	-K^\alpha_j K^k_{\alpha;i} \,,
\label{eqn:VDW-connection}
\end{equation}
where semicolons denote covariant derivatives w.r.t.\ the field space LC connection \eqref{eqn:Levi-Civita}.
In contrast to \eqref{eqn:NDConnection}, the Vilkovisky-DeWitt connection is highly non-local, containing
infinitely many differential operators \cite{Parker:2009}. Based on this connection it is possible to construct
a reparametrization invariant and gauge independent effective action.

To sum up, we discussed three different connections on field space $\config$, all of which have the form given
by equation \eqref{eqn:Christoffels}, using different choices for $A_{ij}^k$. Setting $A_{ij}^k=0$ yields the LC
connection induced by the DeWitt metric, where associated geodesics were calculated in
\cite{DeWitt:1967,Freed:1989,Gil-Medrano:1991}. Choosing relation \eqref{eqn:VDW-connection} gives rise to the
Vilkovisky-DeWitt connection which takes into account the principal bundle character of field space with the gauge
group as structure group. Instead, the choice \eqref{eqn:AForND} leads to connection \eqref{eqn:NDConnection} which
entails the easy exponential parametrization of geodesics. Furthermore, the latter choice is adapted to the
geometric structure of $\mat$, i.e.\ to the local appearance of all metrics in $\config$ as symmetric matrices
with a prescribed signature. This is worked out explicitly in the next section.

\section{Classification of the connection and its geodesics}
\label{sec:GroupTheory}
In this section we describe our results concerning the connection and the corresponding exponential map in terms of
a more general group theory and differential geometry language. It turns out that the connection derived in section
\ref{sec:DerivationNDConnection} is not merely a choice adapted to one particular parametrization but rather
has a more fundamental justification as it arises in a canonical way from the geometry of the space of metrics.
The arguments presented in subsection \ref{sec:GenDesc} are well known, see for instance references
\cite{ONeill:1983,Kobayashi:1969} (cf.\ also \cite{DeWitt:1967},\cite{Freed:1989} and \cite{Percacci:2015wwa}).
They are intended to reconcile the mathematical with the physical literature. Thus, the experienced reader may skip
subsection \ref{sec:GenDesc}.
Here we cover both Euclidean and Lorentzian spacetime metrics at the same time. A distinction becomes necessary only when
studying the global properties of configuration space $\config$; the most important differences will be discussed in
subsection \ref{sec:EuLor}.

\subsection{General description}
\label{sec:GenDesc}
As argued in section \ref{sec:DerivationNDConnection}, a spacetime metric $g\in\config$ at any given spacetime point can
be considered an element of $\mat$ given by \eqref{eqn:DefMatrices}, i.e.\ an element of the space of symmetric matrices
with signature $(p,q)$. Due to this property it is convenient to think of $\config$ as the topological product
$\prod_{x\in M} \mat$, although it is defined more precisely as the space of sections of a fiber bundle with base space
$M$ and typical fiber $\mat$ \cite{Ebin:1970,Freed:1989}. Note that
the arguments presented in this subsection are valid for all $p,q\ge 0$ satisfying $p+q=d$. We observe that for the
search of a geodesic in $\config$ connecting two different metrics $g$ and $g'$ it is sufficient to find a
geodesic in $\mat$ that connects $g_\mn(x)$ to $g'_\mn(x)$ for some $x\in M$ and repeat the construction for all
points in $M$. In that sense the spacetime dependence is trivial since the analysis can be done pointwise
(cf.\ \cite{Freed:1989}).
This notion is compatible with a connection on $\config$ that is ultra-local and diagonal in $x$-space (i.e.\
proportional to $\delta(x-y)\delta(x-z)$), a property that is satisfied by our connection \eqref{eqn:NDConnection} in
particular. For such connections we can reduce our discussion to the matrix space $\mat$ instead of considering
$\config$. Once we have found a geodesic in $\mat$ parametrized by a tangent vector, we obtain a geodesic in $\config$
by using the same parametrization but promoting the tangent vector to an $x$-dependent field. Continuity of the
geodesic with respect to $x$ is then ensured by continuity of the vector field.

We find that $\mat$ is a smooth manifold since it is an open subset in the vector space of all symmetric matrices,
\begin{equation}
 S_d \equiv \left\{A\in\matrices\big|A^T=A\right\}.
\end{equation}
Hence, the tangent space at any point $o\in\mat$ is given by $T_o\mat=S_d$. Here we aim at describing $\mat$ as a
\emph{homogeneous space}. For this purpose we recognize that the group $G\equiv\GLd$ acts transitively on $\mat$ by
\begin{equation}
\begin{split}
	\phi:G\times\mat &\rightarrow\mat, \\
	(g,o) &\mapsto \phi(g,o)\equiv g*o \equiv (g^{-1})^T o\mku g^{-1}.
\end{split}
\label{eqn:groupAction}
\end{equation}
The fact that $g*o$ belongs indeed to $\mat$ and that the action is transitive
(i.e.\ $\forall\ o_1,o_2\in\mat\; \exists\ g\in G: g*o_1=o_2$) is a consequence of Sylvester's law of inertia.
Note that $\phi$ is a left action, that is, $g_1*(g_2*o)=(g_1g_2)*o$.
Let us consider a fixed but arbitrary base point $\bo\in\mat$ now. It is most convenient to think of $\bo$ as
\begin{equation}
I_{p,q}= \begin{pmatrix}\mathds{1}_{p\times p} & \\ & -\mathds{1}_{q\times q}\end{pmatrix},
\label{eqn:Ipq}
\end{equation}
although the subsequent construction is independent of that choice. The \emph{isotropy group} (stabilizer) of $\bo$
is given by\footnote{Note that $h^T\bo\, h=\bo$ is equivalent to $h*\bo \equiv (h^{-1})^T\bo\, h^{-1}=\bo$.}
\begin{equation}
H \equiv H_\bo \equiv \Opq_\bo(p,q)\equiv \left\{ h\in\matrices\big|\, h^T\bo\mku h=\bo\right\},
\end{equation}
which is conjugate to the semi-orthogonal group, and which is a closed subgroup of $G$.
This makes $\mat$ a homogeneous space, and we can write
\begin{equation}
	\mat \simeq G/H,
\end{equation}
where $G/H$ are the \emph{left} cosets of $H$ in $G$.
Defining the \emph{canonical projection}
\begin{equation}
	\pi: G \rightarrow \mat,\; g\mapsto \pi(g)\equiv (g^{-1})^T\bo\mku g^{-1},
\label{eqn:CanProj}
\end{equation}
we see that $(G,\pi,\mat,H)$ becomes a principal bundle with structure group $H$.

Before setting up a connection on the principal bundle let us briefly illustrate the geometric notion behind
this construction. Consider $d$ linearly independent vectors in $\mathbb{R}^d$. This frame can be represented
as a matrix $B\in\GLd$. Now we \emph{fix} a metric $\eta$ by \emph{declaring} the frame to be orthonormal:
\begin{equation}
	\eta(B_{(i)},B_{(j)}) \equiv \delta^{(p,q)}_{ij} \equiv (I_{p,q})_{ij}\, ,
\label{eqn:fixEta}
\end{equation}
where $B_{(i)}$ denotes the $i$-th column of $B$, and $I_{p,q}$ is given by \eqref{eqn:Ipq}. Writing \eqref{eqn:fixEta}
in matrix notation and solving for $\eta$ yields
\begin{equation}
	\eta = (B^{-1})^T I_{p,q}(B^{-1}),
\label{eqn:fixEtaMat}
\end{equation}
so $\eta$ is indeed determined by $B$. We see, however, that the RHS of equation \eqref{eqn:fixEtaMat} is invariant
under multiplications of the type $B\rightarrow B\mku O^{-1}$, where $O\in\Opq(p,q)=\{A\in\matrices|A^T I_{p,q}A=I_{p,q}\}$.
Thus, two frames that differ by a semi-orthogonal transformation define the same metric, so the set of all metrics is
given by $\GLd/\Opq(p,q)$.

In order to find a connection on $(G,\pi,\mat,H)$ we consider the corresponding Lie algebras. In the following, Lie
brackets are given by the commutator of matrices. The Lie algebra $\fg$ of $G$ is the space of all matrices,
\begin{equation}
	\fg=\matrices .
\end{equation}
The Lie algebra of $H$ is the space of ``$\bo$-antisymmetric'' matrices,
\begin{equation}
	\fh = \left\{A\in\matrices\big|\;A^T\bo=-\bo A\right\}.
\label{eqn:defLieh}
\end{equation}
By $\Ad:G\rightarrow\mathrm{Aut}(\fg)$ we denote the adjoint representation of the group $G$,
\begin{equation}
	\Ad(g)(X)=gXg^{-1}\,,\quad g\in G,\, X\in\fg.
\label{eqn:AdRep}
\end{equation}
We find that its restriction $\Ad(H)$ keeps $\fh$ invariant, i.e.
\begin{equation}
	\Ad(h)(\fh) = \fh\quad \forall\, h\in H.
\end{equation}
Let us further define $\fm$ as the space of ``$\bo$-symmetric'' matrices,
\begin{equation}
	\fm \equiv \left\{A\in\matrices\big|\;A^T\bo=\bo A\right\}.
\end{equation}
This defines a vector space complement of $\fh$ in $\fg$,
\begin{equation}
	\fg = \fm \oplus \fh,
\end{equation}
and $\fm$ is called \emph{Lie subspace} for $G/H$. (Note, however, that $\fm$ is not a Lie algebra since
$[m_1,m_2]\in \fh\quad \forall\, m_1,m_2 \in \fm$.) It is easy to show that $\fm$ is invariant under $\Ad(H)$, too,
\begin{equation}
	\Ad(h)(\fm) = \fm\quad \forall\, h\in H.
\end{equation}
Therefore, the homogeneous space $G/H$ is \emph{reductive}.

We use the differential of the canonical projection at the identity $e$ in $G$ in order to make the transition from
the Lie algebra $\fg$ to the tangent space of $\mat$ at $\bo=\pi(e)$,
\begin{equation}
	\diff\pi_e: T_e G\equiv \fg \rightarrow T_\bo\mat.
\end{equation}
Since $\diff\pi_e$ is surjective and has kernel $\fh$, the restriction $\diff\pi_e|_\fm$ is an isomorphism on
the complement $\fm$. Thus, we can \emph{identify} $\fm$ with $T_\bo\mat$.

By means of the left translations $L_g:G\rightarrow G$ we can push forward the Lie subspace $\fm$ to any point $g$
in order to define a distribution on $G$, namely the \emph{horizontal distribution}
\begin{equation}
	\mathcal{H}_g=\diff L_g \fm.
\end{equation}
This defines a \emph{connection} on the principal bundle since it is invariant under the right
translations of $H$:
\begin{equation}
\begin{split}
	\diff R_h(\mathcal{H}_g)&=\diff R_h \diff L_g \fm =\diff L_g \diff R_h \fm =\diff L_g \diff L_h \Ad(h^{-1}) \fm \\
	&= \diff L_g \diff L_h \fm = \diff L_{gh} \fm = \mathcal{H}_{gh}.
\end{split}
\end{equation}
It is called the \emph{canonical connection} of the principal bundle $(G,\pi,\mat,H)$.

The canonical connection, in turn, \emph{induces a connection on the tangent bundle} $T\mat$ which is associated to the
principal bundle \cite{Kobayashi:1969},\footnote{Eq.\ \eqref{eq:DefTM} comprises an implicit reduction of the frame
bundle: Generically the tangent bundle is associated to the frame bundle, $\operatorname{GL}(\mat)$, according to
$T\mat\simeq\operatorname{GL}(\mat)\times_{\operatorname{GL}(D)}\mathbb{R}^D$, where $D\equiv\text{dim}(\mat)=
\frac{1}{2}d(d+1)$. Since the adjoint representation \eqref{eqn:AdRep} maps $H$ to $\operatorname{GL}(D)$ (up to an
isomorphism) and since it is possible to find a principal bundle homomorphism $G\rightarrow\operatorname{GL}
(\mat)$ (with $\mat$ as common base space) compatible with the $H$-action, the structure group is reduced and we have
$\operatorname{GL}(\mat)\times_{\operatorname{GL}(D)}\mathbb{R}^D \simeq G \times_H \fm$.}
\begin{equation}
T\mat \simeq G \times_H \fm \equiv (G \times \fm)/H \,,
\label{eq:DefTM}
\end{equation}
where $h\in H$ acts on $G \times \fm$ by $(g,X)\mapsto (gh^{-1},\Ad(h)X)$. This is often referred to as the
\emph{canonical linear connection} of the homogeneous space $\mat \simeq G/H$.
As we will see below, it can be derived from a metric on $\mat$. In the following we use only the term ``canonical
connection'' since it is clear from the context whether a connection on the principal bundle or on the tangent
bundle is meant.

In general, the torsion tensor following from the canonical connection is given by $T(X,Y)=-\mathrm{pr}_\fm ([X,Y])$ for
$X,Y \in\fm$, where $\mathrm{pr}_\fm$ denotes the projection onto $\fm$ (see e.g.\ reference \cite{Kobayashi:1969}). Here,
since $[\fm,\fm]\subset\fh$, the connection is \emph{torsion free}.

Furthermore, it is possible to define a $G$-\emph{invariant metric on} $\mat$, denoted by $\gamma$. For any
$X,Y\in T_\bo \mat=S_d$ we set
\begin{equation}
  \gamma_\bo(X,Y) \equiv \tr(\bo^{-1}X \mku\bo^{-1}\mku Y) + \frac{c}{2} \tr(\bo^{-1}X)\tr(\bo^{-1}\mku Y),
\label{eqn:metricOnMat}
\end{equation}
with an arbitrary constant $c$. Here, $G$-invariance means that the group action \eqref{eqn:groupAction} of $G$ on $\mat$,
$\phi_g(o)\equiv \phi(g,o)=(g^{-1})^T o\mku g^{-1}$, is \emph{isometric} with respect to this metric:
Since $(\diff\phi_g)_\bo X = (g^{-1})^T X\mku g^{-1}$, we have
\begin{equation}
  \gamma_{\phi_g(\bo)} \big(\mku (\diff\phi_g)_\bo X ,\mku (\diff\phi_g)_\bo Y \mku \big) = \gamma_\bo(X,Y)
\label{eqn:GInvMet}
\end{equation}
for all $X,Y\in T_\bo \mat$. In combination with the $G$-invariance of the canonical connection (w.r.t.\ left translations),
equation \eqref{eqn:GInvMet} has the consequence that the covariant derivative obtained from the canonical connection
\emph{preserves the metric} \eqref{eqn:metricOnMat}. Thus, we conclude that \emph{the canonical connection is the Levi-Civita
connection} on $T\mat$ with respect to $\gamma$ \cite{Kobayashi:1969}.

We can deduce the Levi-Civita connection from \eqref{eqn:metricOnMat}. For $X,Y\in T_\bo \mat$ it is given by
\begin{equation}
 \Gamma_\bo(X,Y) = -\frac{1}{2}\big(X\bo^{-1}Y+Y\bo^{-1}X\big).
\label{eqn:LCOnMInMatrixForm}
\end{equation}

For the sake of completeness we mention that for any point $\bo\in\mat$ there is a symmetry $s_\bo$, i.e.\ a map
$s_\bo:\mat\rightarrow\mat$ which is an element of the isometry group of the metric $\gamma$ and which has the reflection
properties, $s_\bo(\bo)=\bo$ and $(\diff s_\bo)_\bo= -\text{Id}$. It is given by the involution $s_\bo(o)\equiv\bo\mku o^{-1}\bo$
and makes $\mat$ a symmetric space.

With the above groundwork it is straightforward to construct geodesics through the point $\bo$. For that purpose we have to find the
\emph{exponential map} on the manifold $\mat$ with base point $\bo$, here denoted by $\exp_\bo$. On the matrix Lie group $G$ the
exponential map is given by the standard matrix exponential, $\exp$, where we also write $\exp A=\e^A$. As shown in references
\cite{ONeill:1983,Kobayashi:1969}, the map $\exp_\bo \circ\, \diff\pi_e:\fm\rightarrow\mat$ is a local diffeomorphism, and it holds
\begin{equation}
 \exp_\bo \circ\, \diff\pi_e = \pi \circ\, \exp \,.
\end{equation}
Hence, geodesics on $\mat$ are determined by
\begin{equation}
 \exp_\bo X = \pi\big( \e^{\diff\pi_e^{-1}X}\big),
\end{equation}
for $X\in T_\bo\mat=S_d$. From equation \eqref{eqn:CanProj} we obtain $\diff\pi_e^{-1} X = -\frac{1}{2} \bo^{-1} X$,
resulting in
\begin{equation}
\begin{split}
 \exp_\bo X &= \pi\big( \e^{-\frac{1}{2} \bo^{-1} X}\big)
  = \big( \e^{\frac{1}{2} \bo^{-1} X}\big)^T \bo \; \e^{\frac{1}{2} \bo^{-1} X} \\
  &= \bo\, \e^{\bo^{-1} X}.
\end{split}
\label{eqn:GeodesicsParametrization}
\end{equation}
With the identifications $\bo=\bg(x)$ and $X=h(x)$ that is precisely our parametrization \eqref{eqn:ExpParamMatrix} of the
metric.\footnote{This is to be contrasted with the geodesics found in reference \cite{Freed:1989} (see also
\cite{DeWitt:1967}) which are based on the LC connection induced by the DeWitt metric in $\config$. This is equivalent to
determining geodesics in $\mat$ with respect to the LC connection of the metric $\sg\,\gamma$, i.e.\ of our metric
\eqref{eqn:NDMetric} times $\sg$. The resulting parametrization of geodesics has a more involved form than
\eqref{eqn:GeodesicsParametrization}. In the referenced calculations, the authors decompose $\mat$ into a product of
$\mat_\mu$ and $\mathbb{R}^+$, where $\mat_\mu$ are all elements of $\mat$ with determinant $\mu$. Remarkably, geodesics in
$\mat_\mu$ based on $\sg\,\gamma$ have the same structure as our result \eqref{eqn:GeodesicsParametrization} that describes
geodesics in $\mat$ based on $\gamma$. Related to our discussion in section \ref{sec:CompConn}, this can be traced back to
the factor $\sg$ again which is constant in $\mat_\mu$.}
This is the main result of this section. \emph{The exponential parametrization describes geodesics with respect to the
canonical connection}.

To sum up, we have seen that the canonical connection arises in a very straightforward way from the basic fiber bundle
structure of $\mat\simeq G/H$. Since this leads directly to the exponential parametrization, we consider it the most natural
approach to parametrizing metrics.

Finally, we convince ourselves that the metric $\gamma$ on $\mat$ defined in \eqref{eqn:metricOnMat} is identical to
\eqref{eqn:NDMetric}. Setting $\bo=\bg$ and symmetrizing adequately we obtain
\begin{equation}
\begin{split}
 \gamma_{\bg}(X,Y)&=\tr(\bg^{-1}X \bg^{-1}Y) + \frac{c}{2} \tr(\bg^{-1}X) \tr(\bg^{-1}Y) \\
  &= \left( \bg^{\mu(\rho} \bg^{\sigma)\nu} + \frac{c}{2}\, \bg^\mn \bg^{\rho\sigma} \right) X_\mn Y_{\rho\sigma} \\
  &\stackrel{!}{=} \gamma^{\mu\nu\rho\sigma}X_\mn Y_{\rho\sigma}.
\end{split}
\end{equation}
Thus, we find indeed $\gamma^{\mu\nu\rho\sigma}=\bg^{\mu(\rho} \bg^{\sigma)\nu} + \frac{c}{2}\, \bg^\mn \bg^{\rho\sigma}$.

Moreover, the corresponding Christoffel symbols follow directly from equation \eqref{eqn:LCOnMInMatrixForm}, yielding the
same result as in equation \eqref{eqn:Levi-CivitaOnM}. We emphasize that they are independent of the parameter $c$.

\subsection{Euclidean vs.\ Lorentzian metrics}
\label{sec:EuLor}

Next, we specify some topological and geometrical properties of $\mat$, defined in equation \eqref{eqn:DefMatrices},
where we have to distinguish between different signatures. In the following, ``for all $p,q$'' refers to ``for all
$p,q\in\mathbb{N}_0$ with $p+q=d$''.

As already stated above, $\mat$ is an \emph{open subset} in the space of symmetric matrices for all $p,q$. Irrespective of the
signature it is \emph{non-compact}.

Furthermore, it is \emph{path-connected} for all $p,q$. (Note that $G=\GLd$ is non-connected, but the subgroup $H$ has
elements in both of the connected components of $G$).

For the special cases $p=d$, $q=0$ (positive definite matrices) and for $p=0$, $q=d$ (negative definite matrices) the
space $\mat$ is also \emph{simply connected} since it is \emph{convex}. In contrast, \emph{when considering mixed signatures,
$\mat$ is not simply connected}.\footnote{This can be proven by means of the long exact homotopy sequence.}

The scalar curvature of $\mat$ is a negative constant: Independent of $p$, $q$ and the metric parameter $c$, it is given by
\begin{equation}
 R_\mat = -\frac{1}{8}d(d-1)(d+2).
\end{equation}

For all values of $p$ and $q$ we find that $\mat$ is \emph{geodesically complete}, i.e.\ every maximal geodesic is defined
on the entire real line $\mathbb{R}$. It can be shown, for instance algebraically, that $\bo\, \e^{\bo^{-1} X}$ stays in
$\mat$ for all $X\in S_d$. In ref.\ \cite{Nink:2014yya} this has been done for positive definite matrices. Along similar
lines it can be proven for all $p,q$. Here, however, an algebraic proof is not necessary since geodesic completeness is
guaranteed by construction: $\mat$ is a homogeneous space and the exponential map is defined on the entire tangent space.

We emphasize that connectedness plus geodesic completeness does \emph{not} imply that, given any two points in $\mat$,
there exists a geodesic connecting these two points. Actually this is the main difference between the cases of positive
and negative definite matrices on the one hand and matrices with signature $p\ge 1$, $q\ge 1$ on the other hand.
\emph{In case} (a), \emph{$p=d$, $q=0$ or $p=0$, $q=d$, any two points in $\mat$ can be connected by a geodesic,
while for case} (b), \emph{i.e.\ for all other signatures, this is generally not possible}. The deeper reason lies in the
(semi-)Riemannian structure of $\mat$.

Let us consider (a) first. In that case $\mat$ has a Riemannian structure provided that $c\ge -\frac{2}{d}$ since the metric
$\gamma$ given by equation \eqref{eqn:metricOnMat} is positive definite: For both $p=d$, $q=0$ and $p=0$, $q=d$ one can
show that
\begin{equation}
 \gamma_\bo(X,X)=\tr\big((\bo^{-1}X)^2\big)+\frac{c}{2}\big(\tr (\bo^{-1}X)\big)^2 >0 ,
\end{equation}
for all $X\in T_\bo\mat=S_d$, $X\neq 0$, and for $c\ge -\frac{2}{d}$. Therefore, the Hopf--Rinow theorem is applicable, and,
as a consequence, \emph{any two points of $\mat$ can be connected by a geodesic}. The exponential map is a \emph{global
diffeomorphism} then. Since we have already seen that the connection is independent of the parameter $c$, the resulting
geodesics do not depend on $c$ either, and thus, the statement of geodesic connectedness remains true even for $c < -\frac{2}{d}$.
Using algebraic methods, it has already been shown in \cite{Nink:2014yya} that any two points of $\mat$ are connected by means of
the exponential parametrization, but with the arguments presented here we know in addition that this parametrization describes a
geodesic.

The situation is different in case (b): For $p\ge 1$, $q\ge 1$ and for all values of $c$ it is easy to check that
$\gamma_\bo(X,X)$ can become both positive and negative, depending on $X$, so $\gamma$ is indefinite and $\mat$ is
semi-Riemannian.\footnote{It is possible to define a different metric when $p\ge 1$, $q\ge 1$ that makes $\mat$ Riemannian.
However, such a metric would not be $G$-invariant, its Levi-Civita connection would not be the canonical
connection, and it would not extend to a covariant metric in field space $\config$. In particular, corresponding geodesics
would not be given by the simple exponential parametrization.}
This means that the Hopf--Rinow theorem is not applicable. It turns out that there are points in $\mat$ that cannot be connected
by a geodesic. Thus, the exponential map is \emph{not surjective}. But even the restriction to its image does not make it a global
diffeomorphism since it is also \emph{not injective}. To see this we discuss two counterexamples for $2\times 2$-matrices,
that is, for $p=1$ and $q=1$.

First, let us consider the base point
\begin{equation}
	\bo=\begin{pmatrix}1&0\\0&-1\end{pmatrix}, \text{ and }
	X=\begin{pmatrix}0&\alpha\\\alpha&0\end{pmatrix} \in T_\bo\mat.
\end{equation}
This gives rise to the exponential map
\begin{equation}
	o = \bo\, \e^{\bo^{-1}X} = \begin{pmatrix}\cos\alpha&\sin\alpha\\\sin\alpha&-\cos\alpha\end{pmatrix},
\end{equation}
which is periodic, and thus not injective.

Second, we try to connect the base point
\begin{equation}
	\bo=\begin{pmatrix}1&0\\0&-1\end{pmatrix} \text{ to another point }
	o =\begin{pmatrix}-2&0\\0&1\end{pmatrix},
\end{equation}
which clearly belongs to $\mat$, too. That means we have to find an $X\in T_\bo\mat=S_d$ that solves
the equation
\begin{equation}
	\bo^{-1}o = \begin{pmatrix}-2&0\\0&-1\end{pmatrix} = \e^{\bo^{-1}X}.
\label{eqn:Counterex2}
\end{equation}
There is an existence theorem \cite{Culver:1966}, however, which states that a real square matrix
has a \emph{real} logarithm if and only if it is non-degenerate and each of its Jordan blocks
belonging to a negative eigenvalue occurs an even number of times. Thus, since the matrix in the
middle of equation \eqref{eqn:Counterex2} has two distinct negative eigenvalues, it does not have
a real logarithm, so there is no $X\in T_\bo\mat$ that solves \eqref{eqn:Counterex2}. This proves
that the exponential map is not surjective for $p=1$ and $q=1$.

Similar counterexamples can be found for higher dimensions. To sum up, for all non-degenerate symmetric
matrices with mixed signature ($p\ge 1$, $q\ge 1$) the exponential map is neither injective nor surjective.

\begin{figure}[tp]
  \centering
  \includegraphics[width=.9\columnwidth]{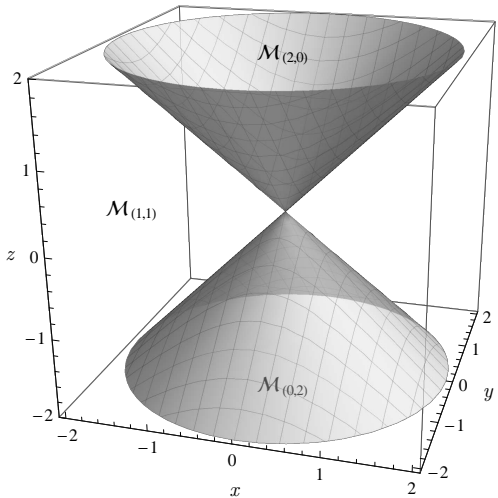}
  \caption{Using parametrization \eqref{eqn:ParamSymMat} the space of symmetric $2\times 2$-matrices decomposes
    into positive definite matrices $\mat_{(2,0)}$ (interior of the cone with positive $z$), negative definite matrices
    $\mat_{(0,2)}$ (interior of the cone with negative $z$), and symmetric matrices with signature $(1,1)$
    ($\mathbb{R}^3$ where the two cones are cut out). The cones extend to $z\rightarrow\pm\infty$. We observe that
    $\mat_{(1,1)}$ is not simply connected.}
  \label{fig:PosDef}
\end{figure}

In the case of $2\times 2$-matrices the space $\mat$ can be illustrated by means of three dimensional plots.
It will turn out convenient to parametrize any symmetric matrix by
\begin{equation}
 \begin{pmatrix} z-x & y \\ y & z+x \end{pmatrix},
\label{eqn:ParamSymMat}
\end{equation}
since the various subspaces assume simple geometric shapes then. The eigenvalues of \eqref{eqn:ParamSymMat} are
given by
\begin{equation}
 \lambda = z \pm \sqrt{x^2+y^2}.
\end{equation}
Thus, the condition for positive definite, negative definite or indefinite matrices, i.e. both eigenvalues positives,
negative or mixed, respectively, leads to a condition for $x$, $y$ and $z$, which can be displayed graphically.
Let $\mat_{(p,q)}$ denote the set of symmetric matrices with signature $(p,q)$. Then the set of all non-degenerate
symmetric $2\times 2$-matrices decomposes into $\mat_{(2,0)}$, $\mat_{(1,1)}$ and $\mat_{(0,2)}$. This is shown in
figure \ref{fig:PosDef}. By use of parametrization \eqref{eqn:ParamSymMat} the set of positive definite matrices,
$\mat_{(2,0)}$, is represented by the inner part of a cone which is upside down and has its apex at the origin. Note
that it extends to $z\rightarrow\infty$. Negative definite matrices, $\mat_{(0,2)}$, are merely a reflection of this
cone through the origin. Finally, $\mat_{(1,1)}$ is mapped to $\mathbb{R}^3$ from which two cones are cut out.
The surface of the cones belongs to neither of the three sets but rather to degenerate symmetric matrices.

\begin{figure}[tp]
  \centering
  \includegraphics[width=.9\columnwidth]{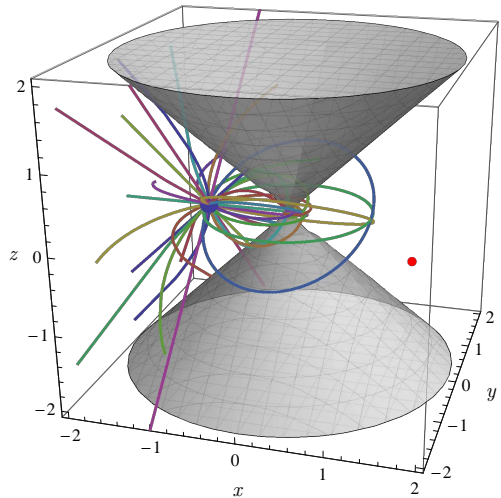}
  \caption{Geodesics in $\mat_{(1,1)}$, starting at $(x,y,z)=(-1,0,0)$, where $\mat_{(1,1)}$ is given by the
    white space without the gray cones. As opposed to the case of positive definite matrices, we find periodic solutions
    here. Moreover, whenever a geodesic traverses the $yz$-plane on the positive $x$ side, it crosses the half-line
    $\{(x,0,0)\in\mathbb{R}^3|x>0\}$. There is no geodesic connecting the base point to the marked point at
    $(x,y,z)=\big(\frac{3}{2},0,-\frac{1}{2}\big)$.}
  \label{fig:Geodesics}
\end{figure}

At last, we illustrate geodesics in $\mat_{(1,1)}$. This helps to understand how it can be possible that every maximal
geodesic is defined on the entire real line, while still not all points can be reached by geodesics starting from a base
point. Figure \ref{fig:Geodesics} shows what happens. By way of example, we choose the base point $\bo\in\mat_{(1,1)}$
with parametrization $(x,y,z)=(-1,0,0)$ and some random tangent vectors that give rise to corresponding geodesics.
We observe that most of the example geodesics lie entirely in the half space with negative $x$. However, those entering
the positive $x$ half space have in common that they run through the same axis: Whenever they cross the $yz$-plane at
positive $x$ they intersect the $x$-axis. This holds for all geodesics starting at $\bo$, that is, at $x>0$ they
can never reach points in the $yz$-plane with $z>0$ or $z<0$. Furthermore, we see the periodic solutions in figure
\ref{fig:Geodesics} as geodesics circling around the origin.

\begin{figure}[tp]
  \centering
  \includegraphics[width=.9\columnwidth]{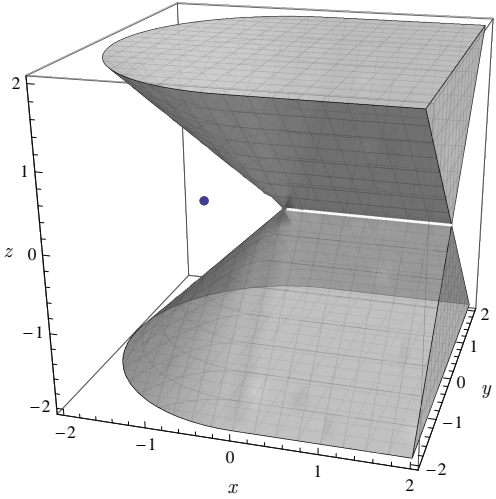}
  \caption{The white region shows the space within $\mat_{(1,1)}$ that can be reached by a geodesic starting from the
    base point at $(x,y,z)=(-1,0,0)$.}
  \label{fig:Reachable}
\end{figure}

By using the existence theorem concerning real logarithms \cite{Culver:1966} it can be shown that the points which
can be reached from the base point by a geodesic are given by the white region in figure \ref{fig:Reachable}.
We find that the two cones effectively shield the space behind them.

In conclusion, the exponential parametrization describes geodesics in the space of metrics, adapted to the fundamental
geometric structure. For Euclidean metrics there is a one-to-one correspondence between tangent vectors and metrics,
while for general/Lorentzian signatures there is not. In the latter case the parametrization can only be cured by
restricting the tangent space and starting from several base points such that all metrics are reached once and only once.


\section{Covariant Taylor Expansions and Nielsen Identities}
\label{sec:Applications}

In the previous sections we have discussed the geometry of the gravitational field space $\config$ in great detail.
It was shown specifically that $\config$ can be equipped with a canonical field space connection \eqref{eqn:NDConnection},
reproducing the exponential parametrization of the metric field. Thus, like any other parametrization based on such a
geodesic formalism, the use of the exponential parametrization allows for the construction of covariant objects, in
particular, of a geometric effective (average) action, which is briefly reviewed in this section. A thorough introduction
to the topic can be found, for instance, in reference \cite{Parker:2009}.

Having a connection $\Gamma^k_{ij}$ on $\config$ at hand, the key idea is to define coordinate charts based on geodesics.
We start by selecting an arbitrary base point $\vpb$ in field space and using $\Gamma^k_{ij}$ to construct geodesics that
connect neighboring points $\vp$ to $\vpb$.\footnote{We assume here that such geodesics exist. This assumption is valid for
Euclidean metrics, but metrics with Lorentzian signatures have to be handled with more care, see section \ref{sec:EuLor}.}
As in section \ref{sec:DerivationNDConnection}, let $\vp^i (s)$ denote such a geodesic connecting
$\vp^i (0) = \vpb^i$ to $\vp^i(1) = \vp^i$. The vector tangent to the geodesic at the starting point $\vpb^i$ is given by
$\frac{\diff \vp^i(s)}{\diff s}\big|_{s=0} = h^i[\vpb,\vp]$. It depends on both base point and end point. We have already
argued that $\config$ is geodesically complete, and that geodesics are determined by the exponential map. Since the
exponential map is a local diffeomorphism, we see that
$\exp_{\vpb} :\operatorname{T}_{\vpb} \config\rightarrow\mathcal{U}\subseteq\config$ with $h \mapsto \vp[h;\vpb]$ constitutes
a coordinate chart. We refer to this chart as geodesic coordinates. Note again that the field $ h^i[\vpb,\vp]$ plays a twofold
role as a tangent vector located at $\vpb$ and as the coordinate representation of the point $\vp$.

On the basis of geodesic coordinates it is possible to perform covariant expansions which can eventually be used to define a
reparametrization invariant effective action. Let $A[\varphi]$ be any scalar functional of the field $\varphi^i$, and let
$\vp^i (s)$ be a geodesic as above. Then the functional $A[\varphi]$ can be expanded as a Taylor series according to
\begin{align}
A[\vp] = A[\vp(1)] = \sum_{n=0}^{\infty} \frac{1}{n!} \left.\frac{\diff^n }{\diff s^n}\right|_{s=0} A[\vp(s)]\,.
\end{align}
By extensively making use of the geodesic equation this relation can be rewritten as \cite{Honerkamp:1972}\cite{Parker:2009}
\begin{align}
\label{eqn:cov_exp}
A [\varphi]
=
\sum_{n=0}^{\infty} \frac{1}{n!}\, A_{i_1\dots i_n}^{(n)}[\bar{\varphi}]\, h^{i_1} \cdots h^{i_n}
\,,
\end{align}
where $A_{i_1\dots i_n}^{(n)} [\vpb] \equiv \calD_{(i_n} \dots \calD_{i_1)} A[\bar{\varphi}]$ denotes the $n$-th covariant
derivative (induced by the field space connection) with respect to $\vp$ evaluated at the base point $\vpb$, and $h^i$ are the
coordinates of the tangent vector $h \in \operatorname{T}_{\vpb} \config$. Relation \eqref{eqn:cov_exp} constitutes a
\emph{covariant expansion of} $A[\vp]$ \emph{in powers of tangent vectors}. Since the field $h^i$ can be thought of as the
coordinate representation of the point $\vp$ when using geodesic coordinates, $\vp=\vp[h;\vpb]$, any scalar functional depends
parametrically on $h$ and on the base point $\vpb$. Let us denote functionals interpreted this way with a tilde, so in geodesic
coordinates we have
\begin{align}
A\big[\vp[h;\vpb]\big] \equiv \tilde{A}[h;\vpb] \,.
\label{eqn:functionalNotation}
\end{align}
Expansion \eqref{eqn:cov_exp} implies a useful relation connecting partial and covariant derivatives which reads
\begin{align}
\label{eqn:relatingDs}
\left.\frac{\delta^n}{\delta h^{i_1}\dots\delta h^{i_n}} \tilde{A}[h;\vpb]\right|_{h=0}
=
\calD_{(i_n} \dots \calD_{i_1)} A[\vpb]\,.
\end{align}
The significance of equation \eqref{eqn:relatingDs} comes from the fact that the right hand side is manifestly covariant, so it
can be used to construct reparametrization invariant objects, while covariance is hidden on the left hand side. Hence, \emph{we
observe that} $\left(\frac{\delta}{\delta h}\right)^n A[\exp_{\vpb}(h)]\big|_{h=0}$ \emph{is covariant}.

Employing the connection \eqref{eqn:NDConnection} with its diagonal character in $x$-space, \emph{a covariant derivative
in field space} $\config$ \emph{reduces to a covariant derivative in target space} $\mat$, which we will denote by
\begin{equation}
\calD_k h^i \equiv \frakD_K h^I \delta(x-y) \equiv \frakD^{\alpha\beta} h_{\mu\nu}(g) \delta(x-y),
\label{eqn:PropertyCovDer}
\end{equation}
where capital Latin labels denote again pairs of spacetime indices, $h^I(x)\equiv h_{\mu\nu}(x)$. Assuming that the functional
$A$ can be written as $A[\varphi] = \int\diff^d x \mathcal{L}(\varphi)$, expansion \eqref{eqn:cov_exp} becomes
\begin{align}
A [\varphi]
=
\int \diff^d x
\sum_{n=0}^{\infty} \frac{1}{n!} \frakD_{(I_n}\dots \frakD_{I_1)} \mathcal{L}[\bar{\varphi}]\;
 h^{I_1}(x) \cdots h^{I_n}(x)
\,.
\end{align}
Thus, with connection \eqref{eqn:NDConnection}, covariant expansions in $\mat$ can be lifted to covariant expansion in
$\config$ \emph{in a minimal way}. Note that, related to our discussion in section \ref{sec:CompConn}, in gravity derivatives
act on the volume element $\sg$ inside $\mathcal{L}$, too, in contrast to the situation in non-linear sigma models.

Let us turn to the quantum theory now. Based on the usual definition, the effective action $\Gamma$ is determined by a
functional integro-differential equation,
\begin{equation}
 \e^{-\Gamma[\vpb]} = \int\mathcal{D}\vp\,\e^{-S[\vp] + (\vp^i-\vpb^i)\frac{\delta\Gamma}{\delta\vpb^i}} \,,
\label{eqn:IntegroDiff}
\end{equation}
where $S$ is the classical action. In the case of gauge theories the functional integral involves an additional integration
over ghost fields, and gauge fixing and ghost action terms are added in the exponent on the RHS. For a discussion of the
measure $\mathcal{D}\vp$ we refer the reader to reference \cite{Mottola:1995}. It is known that $\Gamma$ fails to be
reparametrization invariant. As already noted by Vilkovisky \cite{Vilkovisky:1984st}, the reason for non-covariance  in the
naive definition originates from the source term $(\vp^i-\vpb^i)J_i$ with $J_i=\delta\Gamma/\delta\vpb^i$. Since $\vp^i$
and $\vpb^i$ are merely coordinates, such a term
makes no sense from a geometrical point of view. However, by employing the powerful tools of Riemannian geometry
it is possible to define the path integral covariantly. The key idea is to couple sources to tangent vectors which are
determined by geodesics from $\vpb$ to $\vp$. That means, the source term in \eqref{eqn:IntegroDiff} must be of the form
$S_\mathrm{source} = J^i G_{ij} h^i\equiv J^i G_{ij}[\vpb] h^i[\vpb,\vp]$, where both source field $J^i$ and
fluctuation field $h^i$ are now elements of $\operatorname{T}_{\gb} \config$ over some arbitrary base point $\gb$.
Moreover, the field space metric can be used to include the volume factor $\sqrt{\det G_{ij}}$ in the functional
integral such that the combination $\mathcal{D}\vp \sqrt{\det G_{ij}[\vp]}$ and its analog in terms of $\mathcal{D}h$ are
manifestly covariant. This procedure allows for the construction of a reparametrization invariant effective action
\cite{Vilkovisky:1984st}, referred to as the \emph{geometric effective action}.

Here, we would like to review some properties of the geometric effective action $\Gamma$ and its generalization to the
geometric effective average action $\Gamma_k$ which takes into account scale dependence according to the renormalization
group. We emphasize that the following statements are not restricted to a particular connection, say, the Vilkovisky-DeWitt
connection, but they are \emph{valid for any field space connection}, in particular for the one given by equation
\eqref{eqn:NDConnection}.

The geometric effective action $\Gamma[\vp,\vpb]\equiv\tilde{\Gamma}[h;\vpb]$ in a Euclidean quantum field theory
satisfies the $\hbar$-expansion
\begin{align}
\tilde{\Gamma} [h;\vpb]
=
\tilde{S}[h;\vpb] + \frac{\hbar}{2}\Tr \log \tilde{S}^{(2)}[h;\vpb] + \mathcal{O}(\hbar^2)\,,
\label{eqn:GammaExpansion}
\end{align}
where $\tilde{S}^{(2)}_{ij}[h;\vpb] = \tfrac{\delta^2 \tilde{S}[h;\vpb]}{\delta h^j \delta h^i}$. By adding an infrared
cutoff term $-\frac{1}{2}h^i(\mathcal{R}_k[\vpb])_{ij}\,h^j$ with scale $k$ in the exponent on the RHS of
\eqref{eqn:IntegroDiff}, it is possible to construct a generalization of the geometric $\Gamma$, denoted by
$\Gamma_k$, which is referred to as geometric effective average action \cite{Pawlowski:2003sk,Donkin:2012ud}. Its
running is governed by the functional RG (renormalization group), leading to the flow equation
\cite{Wetterich:1993,Pawlowski:2003sk}
\begin{align}
\del_k \tilde{\Gamma}_k[h;\vpb] = \frac{1}{2} \Tr \Big[
\big(\tilde{\Gamma}_k^{(2)}[h;\vpb] + R_k\big)^{-1}
\del_k R_k \Big] \,.
\label{eqn:FRGE}
\end{align}
Both in \eqref{eqn:GammaExpansion} and in \eqref{eqn:FRGE} the effective (average) action depends additionally on the base
point $\vpb$. In general, an extra $\vpb$-dependence also remains when switching from geodesic coordinates to a $\vp$-based
coordinate chart, $\tilde{\Gamma}_k[ h;\vpb ] = \Gamma_k[\vp,\vpb]$. This extra dependence stems from gauge fixing and cutoff
terms. A single field effective (average) action is usually obtained by taking the coincidence limit $\vpb \to \vp$, or
equivalently, $h \to 0$.

In practice, flows of the effective average action are computed by resorting to the method of truncations, i.e.\ by
constructing $\tilde{\Gamma}_k[h;\vpb]$ out of a restricted set of possible invariants. Most studies based on the functional
RG deal with single field truncations, where the effective average action is approximated by functionals of the form
$\tilde{\Gamma}_k[h;\vpb] = \Gamma_k [\vp(h;\vpb)]$ without extra $\vpb$-dependence. In this case, after taking the field
coincidence limit we can make use of relation \eqref{eqn:relatingDs} on the right hand side of \eqref{eqn:FRGE}, where
we write
\begin{align}
\frac{\delta^2 \tilde{\Gamma}_k[h;\vpb]}{\delta h^i \delta h^j} \bigg|_{h=0}
=
\calD_{(i}\calD_{j)} \Gamma_k [\vpb]
\,.
\end{align}
Thus, we obtain a covariant expression. In particular, this applies to the use of the exponential parametrization: By means
of equation \eqref{eqn:PhiExpansion} we can expand $g=\bg\,\e^{\bg^{-1}h}$ inside $\Gamma_k$ in terms of $h$, that is,
schematically we have
$\Gamma_k\big[\bg\,\e^{\bg^{-1}h},\bg\big]=\Gamma_k\big[\bg+h-\frac{1}{2}\bar{\Gamma}\mku hh+\mO(h^3),\bg\big]$.
Thanks to the appearance of the connection, \emph{a subsequent expansion of} $\Gamma_k$ \emph{in terms of} $h$ \emph{is
covariant}, in contrast to an expansion of $\Gamma[\bg+h,\bg]$ with the linear split \eqref{eqn:StdParam}. This is a very
important property of the exponential parametrization. At second order we have, in uncondensed notation,
\begin{align}
\left.\frac{\delta^2 \Gamma_k[\gb \mku \e^{\bg^{-1}h},\gb ]}{\delta h_{\mu\nu}(x) \delta h_{\alpha\beta}(y)}\right|_{h=0}
=
\calD^{\mu\nu}_{(x)} \calD^{\alpha\beta}_{(y)} \Gamma_k[g,\gb]\Big|_{g=\bg} \;,
\end{align}
where the covariant derivatives act on the first argument of the effective average action, and symmetrization is ensured
by connection \eqref{eqn:NDConnection}.

Above we have mentioned the extra $\vpb$-dependence of the effective (average) action. However, $\tilde{\Gamma}[h;\gb]$
only seemingly depends on two fields. As it has been discussed in
\cite{Burgess:1987,Branchina:2003,Pawlowski:2003sk,Manrique:2009uh,Bridle:2013sra,Becker:2014qya}, it rather depends on a
certain combination of the two fields $g$ and $\gb$, for $\tilde{\Gamma}[h;\gb]$ has to satisfy the generalized Nielsen or
split-Ward identities
\begin{align}
\label{eqn:NielsenID}
\frac{\delta \tilde{\Gamma}}{\delta \bar{\vp}^i} 
  + \braket{\bar{\mathcal{D}}_i \hat{h}^j}\frac{\delta \tilde{\Gamma}}{\delta h^j} = 0 \,,
\end{align}
in the case of non-gauge theories. The tangent vector $\hat{h}^j$ appearing inside the expectation value corresponds to the
integration variable $\hat{\vp}$, i.e.\ we have $\hat{h}^j \equiv \hat{h}^j[\vpb,\hat{\vp}]$. The barred covariant derivative
in \eqref{eqn:NielsenID} acts on the base point,
$\bar{\mathcal{D}}_i \hat{h}^j[\vpb,\hat{\vp}] = \tfrac{\delta \hat{h}^j}{\delta \vpb^i} + \Gamma^j_{ik}[\vpb] \hat{h}^k$.
Relation \eqref{eqn:NielsenID} implies that $\vpb^i$ \emph{and} $h^i$ \emph{can simultaneously be varied in such a way that}
$\tilde{\Gamma}[h;\vpb]$ \emph{is left unchanged}. This is particularly important, as it guarantees that the effective action,
and consequently, all physical quantities, are \emph{independent of the choice of the base point}.
In flat field space $\config$ and in Cartesian coordinates we have $\hat{h}^i[\vpb,\hat{\vp}] = \hat{\vp}^i - \vpb^i$ and thus
$\braket{\bar{\mathcal{D}}_i \hat{h}^j} = - \delta^j_i$. In this special case, relation \eqref{eqn:NielsenID} reduces to the
simple identity
\begin{align}
\frac{\delta \tilde{\Gamma}}{\delta \vpb^i} =\frac{\delta \tilde{\Gamma}}{\delta h^j} \,,
\end{align}
implying a linear split, $\tilde{\Gamma}[h;\vpb] = \Gamma[\vpb + h] = \Gamma[\vp]$.
For gauge theories there are additional terms on the right hand side of \eqref{eqn:NielsenID} due to ghosts and gauge fixing
if a general field space connection different from Vilkovisky-DeWitt is underlying: In this case the zero in
\eqref{eqn:NielsenID} has to be replaced with
\begin{equation}
 \left\langle \frac{\delta S_\text{gf}}{\delta \bar{\vp}^i} \right\rangle 
  + \left\langle \frac{\delta S_\text{gh}}{\delta \bar{\vp}^i} \right\rangle \, .
\label{eqn:GfGh}
\end{equation}

The corresponding relation for the effective average action receives further contributions due to the presence of the regulator.
When using the Vilkovisky-DeWitt connection the modified Nielsen identities read \cite{Pawlowski:2003sk}
\begin{align}
\label{eqn:ModNielsenID}
\frac{\delta \tilde{\Gamma}_k}{\delta \vpb^i}+\braket{\bar{\mathcal{D}}_i\hat{h}^j}\frac{\delta \tilde{\Gamma}_k}{\delta h^j} =
\frac{1}{2}
\Tr G_k \frac{\delta R_k}{\delta \vpb^i}
+
\Tr
R_k G_k \frac{\delta \braket{\bar{\mathcal{D}}_i \hat{h}} }{\delta h}
\,,
\end{align}
with the propagator $G_k =\big(\mku\tilde{\Gamma}_k^{(2)}[h;\gb] + R_k\big)^{-1}$. For a general connection the two terms
in \eqref{eqn:GfGh} have to be added on the right hand side of \eqref{eqn:ModNielsenID}. In the limit $k\to 0$ the identity
\eqref{eqn:ModNielsenID} reduces to the standard form \eqref{eqn:NielsenID}. Another instructive limit is $\braket{\bar{
\mathcal{D}}_i \hat{h}^j}\to -\delta_i^j$ which considers flat field space, where the last term in \eqref{eqn:ModNielsenID}
vanishes. Recently, RG flows satisfying Nielsen identities like \eqref{eqn:ModNielsenID} have been studied in
\cite{Pawlowski:2003sk,Donkin:2012ud,Manrique:2009uh,Bridle:2013sra,Becker:2014qya}. It would be interesting to see to what
extent the geometry of field space corresponding to the exponential parametrization with its property
\eqref{eqn:PropertyCovDer} simplifies the Nielsen identities. We postpone this question to future work, but we conclude
by stating that all geometric identities discussed above are valid when using parametrization \eqref{eqn:ExpParam}.

\section{Conclusions}
\label{sec:Conclusion}

When approaching a quantum theory of gravity on the basis of standard quantum field theory methods involving a path
integral it seems inevitable to introduce a background field $\gb_\mn (x)$. Then, fluctuations $h_\mn(x)$ around this
background field are quantized, assuming the role of variables of integration. We have argued that the path integral
should include only proper metrics, i.e.\ non-degenerate metrics with prescribed signature. This requirement is
implemented in a very natural way by choosing an appropriate metric parametrization, where we identified the exponential
parametrization \eqref{eqn:ExpParam} as the most straightforward choice. Its justification resides in the fact that
\emph{it strictly satisfies the non-degeneracy and signature constraint}, and that \emph{it is adapted to the geometry
of field space} $\config$ at a given spacetime point $x$. The fluctuations $h_\mn$ are interpreted as tangent vectors
which parametrize geodesics in $\config$ starting at $\bg_\mn$ by means of $g_\mn = \bg_{\mu\rho} (\e^h )^\rho{}_\nu$.
We explicitly constructed a connection $\Gamma^k_{ij}$ on field space that reproduces \emph{the exponential
parametrization as the Riemannian exponential map} from tangent space to field space. Thereby, we can identify metrics
$g_\mn$ as points connected to $\gb_\mn$ by geodesics.

The ``naturalness'' of the connection and the resulting parametrization originates from the geometric structure of
field space. Locally, metrics at a given point $x$ can be considered as elements of $\mat$, the space of symmetric
matrices with prescribed signature. We have demonstrated that $\mat$ is a \emph{homogeneous space} which can be written
as $\mat \simeq \GLd/\Opq(p,q)$. For the tangent space $\fg$ of $\GLd$ at the identity there is a vector space
decomposition $\fg = \fm \oplus \fh$ that represents the bundle structure of $\GLd\rightarrow\GLd/\Opq(p,q)$, where
$\fh$ is the Lie algebra of $\Opq(p,q)$ and $\fm$ defines the horizontal direction. Pushing forward the space $\fm$
to other points in $\GLd$ gives rise to a connection on the principal bundle, referred to as the \emph{canonical
connection}. As we have shown, \emph{geodesics on} $\mat$ \emph{induced by this connection are parametrized by the
exponential relation} \eqref{eqn:ExpParam}. In that sense, this parametrization arises canonically.

We have seen that the linear split $g_\mn=\bg_\mn+h_\mn$ as it stands is not suitable in respect of the signature
constraint, which any metric has to satisfy. Therefore, when writing
$\mathcal{D}h_\mn$ in a path integral, it seems reasonable to assume that the $h_\mn$'s are tangent vectors which
parametrize metrics by means of the ``natural'' relation $g_\mn = \bg_{\mu\rho} (\e^h )^\rho{}_\nu$. That is, it is
reasonable to assume that \emph{the functional integral measure is simple when using this parametrization}. If one
adopted the point of view that the measure is simple when $h_\mn$ is defined by the linear parametrization, the transition
to the exponential parametrization would require the introduction of a non-trivial Jacobian \cite{Percacci:2015wwa}.

As a brief remark we would like to mention that, owing to the fact that the metric is a map between two manifolds,
gravity shares many properties with non-linear sigma models, e.g.\ the $G/H$-structure as a homogeneous space
\cite{Friedan:1980}. These models play an important role in many branches of physics, in particular in the context of
symmetry breaking. Recently, breaking of spacetime symmetries in gravity has drawn some attention again
\cite{Delacretaz:2014oxa}. There is, however, a significant difference between non-linear sigma models and the geometry
discussed in the present article. Any metric $G_{ij}$ on field space  $\config$ must contain the volume element $\sg$,
which is field dependent in our case while it is a field independent externally prescribed factor in non-linear sigma
models. We have seen that this factor leads to additional terms in the Levi-Civita connection.

Our approach is to be contrasted with the one of Vilkovisky and DeWitt. While the latter takes into account the bundle
structure of field space with respect to the gauge group, we take into account the canonical bundle structure of the
space of symmetric matrices with prescribed signature. The Vilkovisky-DeWitt method is crucial for constructing gauge
independent quantities like a gauge independent effective action. Due to the non-locality of the connection, however,
it is involved to perform explicit calculations and to determine the corresponding geodesics.
Instead, our method does not aim at gauge independence, but it leads to a local connection giving rise to geodesics
which are described by a simple exponential parametrization. Thus, its advantage are considerable simplifications in
particular calculations.
After all, whether the connection derived here or the Vilkovisky-DeWitt
connection should be used depends on the desired application.

We would like to emphasize that there is a difference between Euclidean and Lorentzian metrics. This difference
is particularly important for the gravitational path integral. In the Euclidean case any two metrics can be connected
by a geodesic based on the canonical connection \eqref{eqn:NDConnection}. Thus, we have geodesic connectedness of
field space $\config$. In contrast, this does not hold in the Lorentzian case: in spite of geodesic completeness,
$\config$ does not exhibit geodesic connectedness. There are points that cannot be reached by geodesics from a
fixed base point $\bg_\mn$, and there are periodic geodesics, i.e.\ the cut locus of $\bg_\mn$ is non-empty.
As a consequence, \emph{in the Euclidean case the path integral} $\int\mathcal{D}h_\mn$ \emph{using the exponential
parametrization captures all metrics once and only once}. \emph{For Lorentzian signatures, however, some metrics are
covered more than once and some are not reached at all}. This flaw can be cured by two steps. (i) One should sum over
several background metrics such that any metric can be reached. (ii) The tangent spaces should be restricted such that
each metric is integrated over only once.

Having established a connection between the exponential parametrization and the geometry of field space $\config$,
we have argued that this parametrization is appropriate for the construction of covariant quantities with respect to
the field space connection \eqref{eqn:NDConnection}. The use of the geodesic formalism allows for covariant Taylor
expansions and the definition of a geometric effective (average) action. With regard to bi-metric truncations for
gravity it would be interesting to see if the geometry of field space with the exponential parametrization can
further simplify the Nielsen identities and renormalization group flows. It remains an open question, too, whether
the ideas presented here can be combined with those of Vilkovisky and DeWitt, that is, whether it is possible to find
a simple geometric parametrization which respects to some extent the gauge bundle structure of field space.
Remarkably, at one loop level the exponential parametrization considered here can already be sufficient to ensure
gauge independence \cite{Falls:2015}.

\begin{acknowledgments}
The authors would like to thank Martin Reuter and Omar Zanusso for many helpful discussions.
\end{acknowledgments}


\begin{thebibliography}{99}

\bibitem{Percacci:1991}
  R.~Percacci, Nucl.\ Phys.\ B {\bf 353} (1991) 271.

\bibitem{DeWitt:2003}
  B.~S.~DeWitt,
  {\it The Global Approach to Quantum Field Theory},
  Clarendon Press, Oxford (2003).

\bibitem{Kawai:1993}
  H.~Kawai, Y.~Kitazawa and M.~Ninomiya, Prog.\ Theor.\ Phys.\ Supp.\ \textbf{114} (1993) 149;
  Nucl.\ Phys.\ B \textbf{393} (1993) 280;
  Nucl.\ Phys.\ B \textbf{404} (1993) 684;
  Nucl.\ Phys.\ B \textbf{467} (1996) 313;\\
  T.~Aida, Y.~Kitazawa, H.~Kawai and M.~Ninomiya, Nucl.\ Phys.\ B \textbf{427} (1994) 158;\\
  J.~Nishimura, S.~Tamura, A.~Tsuchiya, Mod.\ Phys.\ Lett.\ A \textbf{9} (1994) 3565;\\
  T.~Aida and Y.~Kitazawa, Nucl.\ Phys.\ B \textbf{491} (1997) 427.

\bibitem{Nink:2014yya}
  A.~Nink,
  Phys.\ Rev.\ D {\bf 91} (2015)  044030.

\bibitem{DeWitt:1967}
  B.~S.~DeWitt,
  Phys.\ Rev.\ {\bf 160} (1967) 1113.

\bibitem{Freed:1989}
  D.~S.~Freed and D.~Groisser,
  Michigan Math.\ J.\ \textbf{36} (1989) 323.

\bibitem{Percacci:2015wwa}
  R.~Percacci and G.~P.~Vacca,
  Eur.\ Phys.\ J.\ C \textbf{75} (2015), 188.

\bibitem{Labus:2015ska}
  P.~Labus, R.~Percacci and G.~P.~Vacca,
  arXiv:1505.05393.

\bibitem{Codello:2014}
  A.~Codello and G.~D'Odorico, Phys.\ Rev.\ D \textbf{92} (2015) 024026.

\bibitem{Eichhorn:2013}
  A.~Eichhorn,
  Class.\ Quant.\ Grav.\ \textbf{30} (2013) 115016;
  JHEP \textbf{1504} (2015) 096.

\bibitem{Falls:2015}
  K.~Falls,
  arXiv:1501.05331;
  arXiv:1503.06233.

\bibitem{David:1988}
  F.~David, Mod.\ Phys.\ Lett.\ A \textbf{3} (1988) 1651;\\
  J.~Distler and H.~Kawai, Nucl.\ Phys.\ B \textbf{321} (1989) 509;\\
  J.~Polchinski, Nucl.\ Phys.\ B \textbf{324} (1989) 123.

\bibitem{Weinberg:1980}
  S.~Weinberg, in: \textit{General Relativity}, S.~W.~Hawking and W.~Israel (Eds.),
  Cambr.\ Univ.\ Press (1980) 790.

\bibitem{Tsao:1977}
  H.-S.~Tsao, Phys.\ Lett.\ B \textbf{68} (1977) 79;\\
  L.~S.~Brown, Phys.\ Rev.\ D \textbf{15} (1977) 1469;\\
  H.~Kawai and M.~Ninomiya, Nucl.\ Phys.\ B \textbf{336} (1990) 115;\\
  I.~Jack and D.~R.~T.~Jones, Nucl.\ Phys.\ B \textbf{358} (1991) 695.

\bibitem{Borchers:1960}
  H.-J.~Borchers, Il Nuovo Cimento \textbf{15}--5 (1960) 784;\\
  S.~R.~Coleman, J.~Wess and B.~Zumino, Phys.\ Rev.\ \textbf{177} (1969) 2239;\\
  R.~E.~Kallosh and I.~V.~Tyutin, Yad.\ Fiz.\ \textbf{17} (1973) 190
  and Sov.\ J.\ Nucl.\ Phys.\ \textbf{17} (1973) 98.

\bibitem{Mottola:1995}
  E.~Mottola, 
  J.\ Math.\ Phys.\ \textbf{36} (1995) 2470.

\bibitem{Vilkovisky:1984st}
  G.~A.~Vilkovisky,
  Nucl.\ Phys.\ B {\bf 234} (1984) 125.

\bibitem{DeWitt:1987}
  B.~S.~DeWitt,
  in: \textit{Quantum Field Theory and Quantum Statistics},
  I.~A.~Batalin, C.~J.~Isham and G.~A.~Vilkovisky (Eds.), Adam Hilger, Bristol (1987).

\bibitem{Burgess:1987}
  C.~P.~Burgess and G.~Kunstatter, Mod.\ Phys.\ Lett.\ A \textbf{2} (1987) 875;\\
  G.~Kunstatter, Class.\ Quant.\ Grav.\ \textbf{9} (1992) S157.

\bibitem{Ebin:1970}
  D.~G.~Ebin, 
  Bull.\ Amer.\ Math.\ Soc.\ \textbf{74} (1968) 1001;
  Proc.\ Symp.\ Pure Math.\ \textbf{15}, AMS (1970) 11.

\bibitem{Gil-Medrano:1991}
  O.~Gil-Medrano and P.~W.~Michor,
  Quart.\ J.\ Math.\ (Oxford) \textbf{42} (1991) 183.

\bibitem{Blair:2000}
  D.~E.~Blair,
  in \textit{Handbook of Differential Geometry}, Vol.~1,
  F.~J.~E.~Dillen and L.~C.~A.~Verstraelen (Eds.), Elsevier Science, Amsterdam (2000).

\bibitem{DeWitt:1965jb}
  B.~S.~DeWitt,
  Conf.\ Proc.\ C {\bf 630701} (1964) 585
   [Les Houches Lect.\ Notes {\bf 13} (1964) 585].

\bibitem{Huggins:1987zw}
  S.~R.~Huggins, G.~Kunstatter, H.~P.~Leivo\\ and D.~J.~Toms,
  Nucl.\ Phys.\ B {\bf 301} (1988) 627.

\bibitem{Friedan:1980}
  D.~H.~Friedan,
  Phys.\ Rev.\ Lett.\ \textbf{45} (1980) 1057;
  Annals Phys.\ \textbf{163} (1985) 318;\\
  P.~S.~Howe, G.~Papadopoulos and K.~S.~Stelle,
  Nucl.\ Phys.\ B \textbf{296} (1988) 26.

\bibitem{Parker:2009}
  L.~E.~Parker and D.~J.~Toms, {\it Quantum field theory in curved spacetime},
  Cambridge University Press, Cambridge, (2009).

\bibitem{ONeill:1983}
  B.~O'Neill,
  {\it Semi-Riemannian Geometry With Applications to Relativity},
  Academic Press, New York (1983).

\bibitem{Kobayashi:1969}
  S.~Kobayashi and K.~Nomizu,
  {\it Foundations of differential geometry}, Vol.\ I \& II, Wiley, New York (1969).

\bibitem{Culver:1966}
  W.~J.~Culver, Proc.\ Am.\ Math.\ Soc.\ {\bf 17}--5 (1966) 1146.

\bibitem{Honerkamp:1972}
  J.~Honerkamp,
  Nucl.\ Phys.\ B \textbf{36} (1972) 130.

\bibitem{Pawlowski:2003sk}
J.~M.~Pawlowski,
arXiv:hep-th/0310018.

\bibitem{Donkin:2012ud}
I.~Donkin and J.~M.~Pawlowski,
arXiv:1203.4207.

\bibitem{Wetterich:1993}
  C.~Wetterich, Phys.\ Lett.\ B \textbf{301} (1993) 90;\\
  T.~R.~Morris, Int.\ J.\ Mod.\ Phys.\ A \textbf{9} (1994) 2411;\\
  M.~Reuter, Phys.\ Rev.\ D \textbf{57} (1998) 971.

\bibitem{Branchina:2003}
  V.~Branchina, K.~A.~Meissner and G.~Veneziano,
  Phys.\ Lett.\ B \textbf{574} (2003) 319.

\bibitem{Manrique:2009uh}
  E.~Manrique and M.~Reuter,
  Annals Phys.\  {\bf 325} (2010) 785;\\
  E.~Manrique, M.~Reuter and F.~Saueressig,
  Annals Phys.\  {\bf 326} (2011) 440;
  Annals Phys.\  {\bf 326} (2011) 463.
 
\bibitem{Bridle:2013sra}
  I.~H.~Bridle, J.~A.~Dietz and T.~R.~Morris,
  JHEP {\bf 03} (2014) 093;\\
  J.~A.~Dietz and T.~R.~Morris,
  JHEP \textbf{1504} (2015) 118.

\bibitem{Becker:2014qya}
  D.~Becker and M.~Reuter,
  Annals Phys.\  {\bf 350} (2014) 225.

\bibitem{Delacretaz:2014oxa}
  L.~V.~Delacr\'{e}taz, S.~Endlich, A.~Monin, R.~Penco and F.~Riva,
  JHEP {\bf 1411} (2014) 008;\\
  Y.~Hidaka, T.~Noumi and G.~Shiu,
  Phys.\ Rev.\ D \textbf{92} (2015) 045020.

\end{thebibliography}
\end{document}